\documentclass[%
 longbibliography,
 reprint,
 amsmath,amssymb,
 aps,
 pra,
 superscriptaddress
]{revtex4-2}

\usepackage{bm}
\usepackage{braket}
\usepackage[utf8]{inputenc}
\usepackage{qcircuit}
\usepackage{color}
\usepackage{comment}
\usepackage{physics}
\usepackage{bbold}
\usepackage{comment}
\usepackage{amsmath, amssymb, amsthm}
\usepackage{mathtools}
\usepackage{graphicx}
\usepackage[normalem]{ulem}
\usepackage[whole]{bxcjkjatype}
\usepackage[breaklinks]{hyperref}

\theoremstyle{plain}

\theoremstyle{definition}

\theoremstyle{remark}

\usepackage{algorithm}
\usepackage{algorithmicx}
\usepackage{algpseudocode}

\newcommand{\argmin}{\mathop{\rm arg~min}\limits}

\algnewcommand{\Initialize}[1]{%
  \State \textbf{initialize:}
  \State \hspace*{\algorithmicindent}\parbox[t]{0.8\linewidth}{\raggedright #1}
}

\begin{document}

\title{Stochastic Gradient Line Bayesian Optimization\\for Efficient Noise-Robust Optimization of Parameterized Quantum Circuits}

\author{Shiro Tamiya}
\email{tamiya@qi.t.u-tokyo.ac.jp}
\affiliation{Department of Applied Physics, Graduate School of Engineering, The University of Tokyo, 7-3-1 Hongo, Bynkyo-ku, Tokyo 113-8656, Japan}
\author{Hayata Yamasaki}
\email{hayata.yamasaki@gmail.com}
\affiliation{Institute for Quantum Optics and Quantum Information -- IQOQI Vienna, Austrian Academy of Sciences, Boltzmanngasse 3, 1090 Vienna, Austria}
\affiliation{Atominstitut,  Technische  Universit{\"a}t Wien, Stadionallee 2, 1020 Vienna, Austria}
 
\begin{abstract}
Optimizing parameterized quantum circuits is a key routine in using near-term quantum devices.
However, the existing algorithms for such optimization require an excessive number of quantum-measurement shots for estimating expectation values of observables and repeating many iterations, whose cost has been a critical obstacle for practical use.
We develop an efficient alternative optimization algorithm, stochastic gradient line Bayesian optimization (SGLBO), to address this problem. 
SGLBO reduces the measurement-shot cost by estimating an appropriate direction of updating circuit parameters based on stochastic gradient descent (SGD) and further utilizing Bayesian optimization (BO) to estimate the optimal step size for each iteration in SGD\@. 
In addition, we formulate an adaptive measurement-shot strategy and introduce a technique of suffix averaging to reduce the effect of statistical and hardware noise. 
Our numerical simulation demonstrates that the SGLBO augmented with these techniques can drastically reduce the measurement-shot cost, improve the accuracy, and make the optimization noise-robust.
\end{abstract}

\maketitle

\section{Introduction}

\begin{figure*}[t]
    \centering
    \includegraphics[width=7.0in]{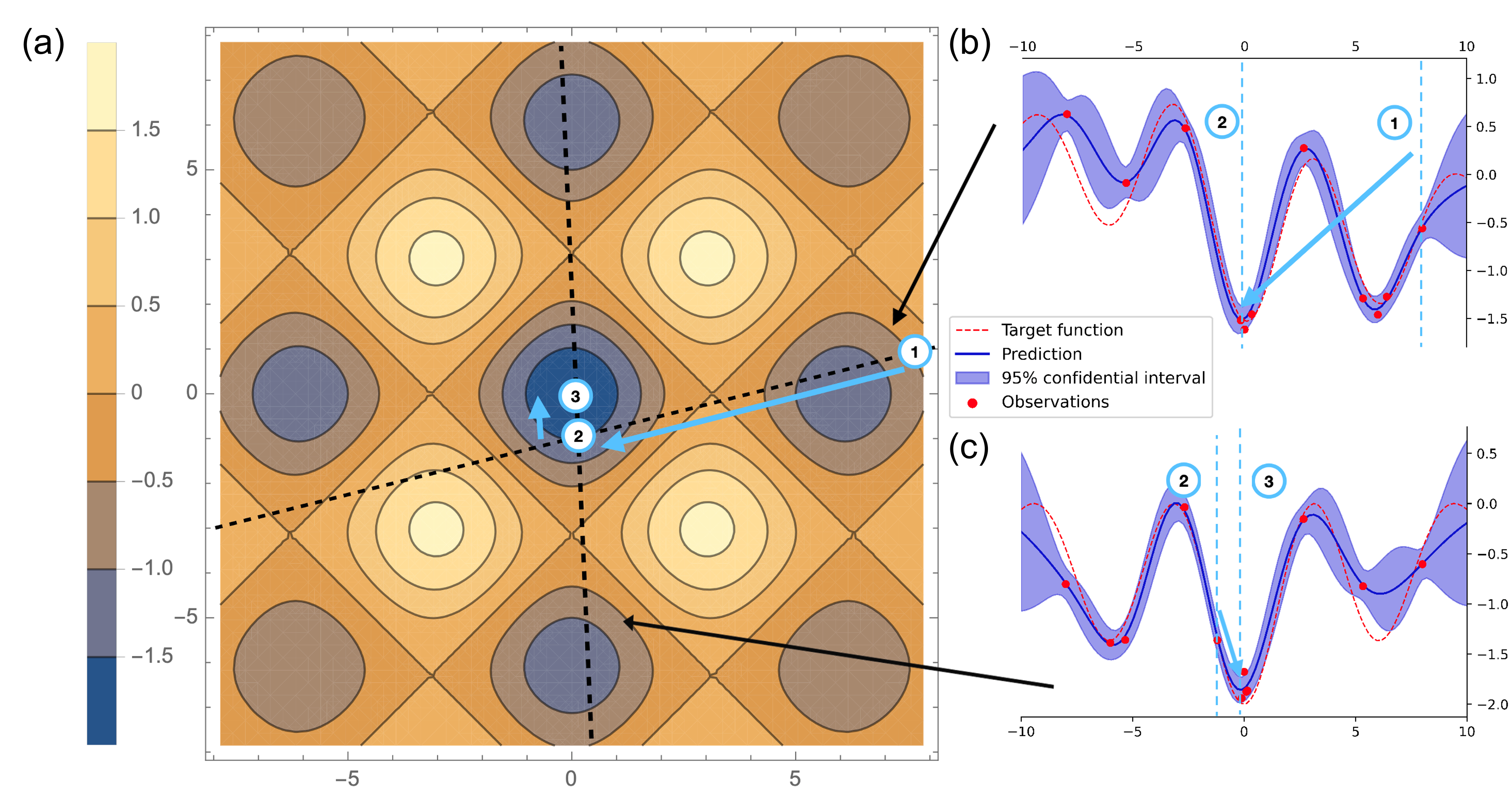}
    \caption{\textbf{An illustration of two iterations in SGLBO for minimizing a $2$D cost function.} (a) The figure  represents the updating procedure of SGLBO on the landscape of the cost function. In particular, in the first iteration, at an initial point 1, we estimate a direction of the gradient of the cost function based on SGD and perform BO on the $1$D subspace in this direction to estimate the optimal step size. (b) Then, we reach point 2 from the point 1 by moving in the estimated direction by the estimated optimal step size. (c) Next, at point 2, we perform the same procedure of estimating the gradient based on the SGD and estimating the optimal step size by the BO on the line of the $1$D subspace, to move from point 2 to point 3. We iterate these procedures until SGLBO converges or consumes a preset number of measurement shots. After all these iterations, SGLBO returns a suffix average over the points visited in the iterations as an output.}
\label{fig: SGLBO}
\end{figure*}

Advances in technologies of quantum hardware lead to intensive research on finding practical applications on noisy intermediate-scale quantum (NISQ) devices~\cite{Preskill2018quantumcomputingin}.
Variational quantum algorithms (VQAs)~\cite{2021_vqa,Endo_2021,bharti2021noisy} are a class of promising candidates of quantum algorithms that are implementable on the NISQ devices.
The VQAs can be used for a variety of computational tasks including quantum chemistry calculations~\cite{Peruzzo2014,Kandala2017,McClean_2016,RevModPhys.92.015003}, combinatorial optimization~\cite{farhi2014quantum,PhysRevX.10.021067,Harrigan_2021}, and training of machine learning models~\cite{Havl_ek_2019,Romero_2017,article_generative,PhysRevLett.122.040504}.
These tasks are achieved by minimizing task-specific cost functions usually defined as a sum of expectation values of observables.
The optimization of minimizing the cost function is performed through updating parameters of a parameterized quantum circuit using a classical optimizer in a feedback loop.
In particular, VQAs employ a quantum device to prepare quantum states that the parameterized quantum circuit outputs.
We perform a shot of quantum measurement on each output state to extract classical information, which is useful for estimating the expectation values of the cost function.
The measurement outcomes are fed to the classical optimizer,
with which we improve the circuit parameters so as to minimize the cost function iteratively.

But problematically, if we try to estimate the expectation values with high precision in the VQAs, we usually need an excessive number of measurement shots until minimizing the cost function~\cite{PhysRevA.92.042303,gonthier2020identifying}.
In practice, a user of a quantum computer often needs to access a distant server of a quantum computer to query measurement shots, while the classical optimizer can be performed locally by the user at a negligible cost compared to the cost of using the quantum computer in terms of time and money~\cite{Sung_2020}; in this setting, the number of measurement shots crucially dominates the cost of VQAs, which we aim to minimize here.
In previous research, problems of reducing computational resources in VQAs have often been tackled by estimating an expectation value efficiently~\cite{Huggins_2021,Huang_2020,2021_derandomized,arrasmith2020operator} and reducing the number of iterations until convergence~\cite{PhysRevResearch.2.043158,Wilson_2021,koczor2021quantum,Ostaszewski_2021,PRXQuantum.2.020329,Stokes_2020,Self_2021, haug2021optimal}.
By contrast, to overcome a dominant obstacle in the above setting of VQAs, we here study the problem of reducing the overall cost of measurement shots in the optimization, that is, how we can optimize the circuit parameters at as little cost of the total number of measurement shots as possible.
A difficulty of this problem stems from the nature of quantum mechanics: it is costly to extract expectation values as classical information from quantum states, yet the optimization would be hard without the assistance of classical information obtained from measurements on the quantum states.
We stress that the problem here is not the estimation of the expectation values themselves; rather, a fundamental question that we ask is how efficiently we can use classical information of the measurement outcomes to optimize the circuit parameters without extracting the expectation values with high precision.

In this work, we address this problem by establishing a framework for the classical optimizer that combines two different optimization approaches, namely, stochastic gradient descent (SGD) and Bayesian optimization (BO).
SGD is a standard algorithm in machine learning for training models, using an estimator of gradient at each optimization step rather than the exact value of the gradient~\cite{10.1214/aoms/1177729586, bottou2018optimization}.
Among a variety of existing optimizers proposed for VQAs~\cite{PhysRevResearch.2.043158,Wilson_2021,koczor2021quantum,Ostaszewski_2021,Sweke2020stochasticgradient,K_bler_2020,gu2021adaptive,Stokes_2020,Lavrijsen_2020,PRXQuantum.2.020329,Self_2021,haug2021optimal}, gradient-based optimizers have been studied intensively, motivated by the fact that the use of gradient information improves convergence~\cite{Harrow_2021}.
Recently, SGD for VQAs has been investigated as a class of gradient-based optimizers~\cite{Sweke2020stochasticgradient}.
The SGD for VQAs often uses a fixed small number of measurement shots to estimate the gradient, which may successfully avoid measuring expectation values with high precision.
However, SGD has major shortcomings that may make the algorithm inefficient. First, instead of the low cost of each iteration, SGD may need a larger number of iteration until convergence than optimization algorithms using the exact gradient; second, SGD requires careful control of the step size of updating the parameters in each iteration, which may crucially affect the efficiency of the algorithm, but an appropriate choice of the step size is often difficult.
On the other hand, BO is another common algorithm for optimization of a black-box function without necessarily using its gradient, which is especially suitable for optimizing imprecise and expensive-to-evaluate functions~\cite{7352306,article_practical_bayesian}.
The BO has many successful applications such as computer vision, robotics, and experimental designs~\cite{10.5555/3042817.3042832,Cantin,Lizotte,DBLP:conf/aaai/AzimiFFBCFLJS10}.
Due to its robustness against noise in the imprecise evaluation of the functions~\cite{7352306,article_practical_bayesian}, BO may also be useful for the optimization in VQAs~\cite{otterbach2017unsupervised,2019_Training}.
However, it is known that BO becomes intractable in high-dimensional settings (typically $\geqq$ 10)~\cite{pmlr-v37-kandasamy15}, and the number of parameters to be optimized in VQAs is usually too large to apply the BO directly.

To retain advantages of SGD and BO in VQAs while compensating for their shortcomings,
we here construct the alternative framework for the optimization of parameterized circuits, stochastic gradient line Bayesian optimization (SGLBO), as illustrated in Fig.~\ref{fig: SGLBO}.
The key idea of SGLBO is that we estimate an appropriate direction of updating the circuit parameters based on SGD, and also utilize BO to estimate the optimal step size in a $1$D direction of the estimated gradient in each iteration.
This idea aims at simultaneously resolving the problems of the step size in the SGD and of the infeasibility of high-dimensional optimization with the BO\@.
To enhance the performance further, we combine the SGLBO with two noise-reducing techniques: adaptive shot strategy and suffix averaging.
The adaptive shot strategy is a technique for dynamically determining the number of measurement shots to be used for the estimation of the gradient~\cite{Friedlander_2012,article_adaptive_sampling,article,article_Pasupathy,pmlr-v54-de17a,balles2017coupling,bollapragada2018a,K_bler_2020}.
We here develop an adaptive shot strategy suitable for SGLBO\@, based on a technique of the norm test~\cite{article,article_adaptive_sampling,pmlr-v54-de17a}.
The norm test combined with SGD is known to provide faster convergence~\cite{pmlr-v54-de17a,article}, and in the case of SGLBO, the norm test reduces not only the number of iterations but also the overall number of measurement shots.
On the other hand,
suffix averaging is a technique for achieving noise reduction.
Instead of directly using the point of the final iteration in the optimization as an estimate of the minimizer of the cost function, the suffix averaging technique uses the average over a latter part of the sequence of points obtained from the iterations~\cite{10.5555/3042573.3042774,harvey2019tight,shamir2013stochastic}. 
We utilize this technique to reduce the statistical noise in estimating the gradient and the optimal step size in SGLBO, and also reduce the effect of the hardware noise of the quantum device\@.

To show the significance of the SGLBO,
we numerically demonstrate that SGLBO can find an estimate of the minimizer of the cost function with a significantly small number of overall measurement shots compared to other state-of-art optimizers~\cite{PhysRevResearch.2.043158,K_bler_2020,kingma2015adam}, in representative tasks for the VQAs, i.e., variational quantum eigensolver~\cite{Peruzzo2014} and variational quantum compiling~\cite{Khatri_2019}.
Thus, the reduction of the number of iterations achieved by finding the optimal step size by BO indeed contributes to the overall reduction of the number of measurement shots.
We also discover that the SGLBO turns out to outperform the state-of-art optimizers not only in terms of the number of measurement shots but also the accuracy in estimating the minimum of the cost functions used in the simulation.
Remarkably, we discover that even under a moderate amount of hardware noise, the SGLBO can estimate the minimum in a task with almost the same accuracy as noiseless cases, whereas the other state-of-the-art optimizers cannot in the same task. 
These results indicate that the SGLBO is a promising approach to reduce the number of measurement shots in the VQAs, and also to make the VQAs more feasible under unavoidable hardware noise in near-term quantum devices.
Note that combination of SGD and BO has been previously studied only in a specific machine-learning setting~\cite{10.5555/3122009.3176863}, but its applicability and advantage for other tasks such as VQAs have been unknown; by contrast, our crucial contribution is to formulate SGLBO as the efficient and noise-robust framework for the task of optimizing parameterized quantum circuits and further develop the techniques of adaptive shot strategy and suffix averaging to demonstrate its advantage in this optimization task.

Consequently, the SGLBO establishes an alternative approach for efficient quantum-circuit optimizers, progressing beyond the existing state-of-the-art optimizers~\cite{PhysRevResearch.2.043158,K_bler_2020,kingma2015adam};
in particular, the novelty of SGLBO is to integrate two different optimization approaches, SGD and BO, to eliminate their shortcomings and take their advantages.
Augmented with the further techniques of adaptive shot strategy and suffix averaging,
the SGLBO is shown to have a significant advantage in the reduction of the cost of the number of measurement shots and also in the robustness against hardware noise, compared to the state-of-the-art optimizers for VQAs\@.
These results open a way to practical algorithm designs for more efficient quantum-circuit optimization in terms of the overall cost of measurement shots, by avoiding both the precise estimation of expectation values and the many iterations of updating circuit parameters; at the same time, the approach developed for the SGLBO provides a fundamental insight into how VQAs can use classical information extracted from quantum states beyond estimating expectation values.

In the rest of this section, we describe the problem setting of optimization tasks in VQAs and review SGD and BO\@.

\begin{figure}
 \includegraphics[width=3.4in]{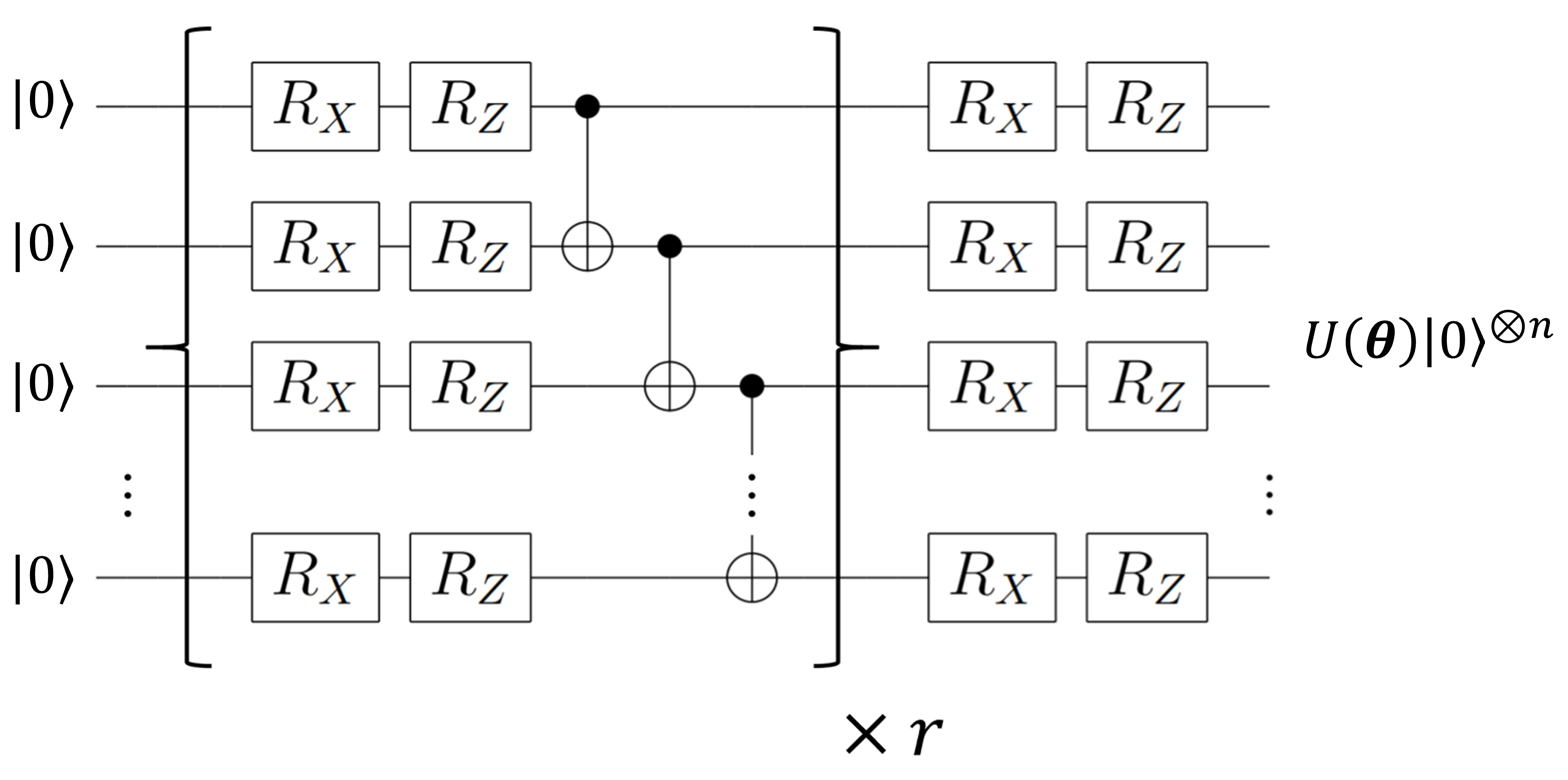}\caption{\textbf{An example of a parameterized quantum circuit used as an ansatz in VQAs.} The circuit parameters $\bm{\theta}=[\theta_1, \cdots, \theta_{D}]^\top \in \mathbb{R}^{D}$ with $D=2n(r+1)$ elements are individually allocated as each rotation angle of Pauli rotation gates $R_X(\theta_i) \coloneqq e^{-\mathrm{i} \theta_i X}$ and $R_Z(\theta_j) \coloneqq e^{-\mathrm{i} \theta_j Z}$. The part of the circuit surrounded by the braces is repeated $r$ times, where the repeated parts may have different parameters.}
\label{fig: ansatz} 
\end{figure}

VQAs~\cite{2021_vqa,Endo_2021,bharti2021noisy} are a class of algorithms that use a parameterized quantum circuit $U(\bm{\theta})$ to minimize a task-specific cost function $f(\bm{\theta})$.
The vector $\bm{\theta}=[\theta_1, \cdots, \theta_{D}]^\top \in \mathbb{R}^{D}$ of $D$ arguments of $f$ is used as the circuit parameters of $U(\bm{\theta})$. 
The cost function $f(\bm{\theta})$ in VQAs is conventionally defined as an expectation value of an observable $O$ on $n$ qubits, with respect to a quantum state output by the parameterized circuit, i.e.,
\begin{equation}
    f(\bm{\theta})=\Tr[O U(\bm{\theta}){(\Ket{0}\Bra{0})}^{\otimes n} U^{\dag}(\bm{\theta})],
    \label{eq: cost function}
\end{equation}
where $\Ket{0}$ is a standard-basis state used for initialization of each qubit, $U(\bm{\theta})\Ket{0}^{\otimes n}$ is the output state of the $n$-qubit parameterized circuit, and $U^{\dag}(\bm{\theta})$ is the complex conjugate of $U(\bm{\theta})$.
The observable $O$ is expanded as a sum of $n$-qubit tensor products of Pauli operators
\begin{equation}
\label{eq:observable}
    O=\sum_{k}c_k P_k,
\end{equation}
where $c_k$ for each $k$ is a real coefficient of the $k$th term, and $P_k$ is a tensor product of $n$ single-qubit Pauli operators
$P_k=\bigotimes_{l=1}^{n}P_{k,l}$
with $P_{k,l} \in \{I, X, Y, Z\}$ being a Pauli (or identity) operator acting on the $l$th qubit.
Here, the identity operator is denoted by $I \coloneqq \ket{0}\bra{0}+ \ket{1}\bra{1}$, and Pauli operators acting on a single qubit are $X\coloneqq \ket{0}\bra{1}+\ket{0}\bra{1}$, $Y\coloneqq -\mathrm{i}\ket{0}\bra{1}+\mathrm{i}\ket{1}\bra{0}$, and $Z\coloneqq \ket{0}\bra{0}-\ket{1}\bra{1}$. 
In a usual setting of VQAs, $U(\bm{\theta})$ is composed of non-parametric gates such as \textsc{CNOT} gates, and parametric gates in the form of \begin{equation}
\label{eq:parametric_gate}
    U(\theta_i)=\exp(-\mathrm{i}P_i\theta_i),
\end{equation}
where $P_i$ is also a tensor product of $n$ single-qubit Pauli operators in the same way as $P_k$ in Eq.~\eqref{eq:observable}. 
For example, Fig.~\ref{fig: ansatz} shows a representative choice of parameterized circuits used for VQAs~\cite{bharti2021noisy}.
Note that the parameter space of the circuit in Fig.~\ref{fig: ansatz} is a $D$-dimensional hypercube $\bm{\theta}\in[-\pi,\pi]^D$, i.e., a bounded subspace of $\mathbb{R}^{D}$, on which a uniform probability distribution is well defined.

The task in the VQAs is to obtain an estimate of the minimum of the cost function
\begin{equation}
    \min_{\bm{\theta} \in \mathbb{R}^{D}}f(\bm{\theta}).
\label{eq: problem statement minimization}
\end{equation}
The minimizer is denoted by
\begin{equation}
    \bm{\theta}^{*} = \argmin_{\bm{\theta} \in \mathbb{R}^{D}}f(\bm{\theta}).
\label{eq: problem statement}
\end{equation}
Note that the cost function $f(\bm{\theta})$, in general, can be non-convex, and it can be computationally hard in general to obtain the exact solution of the optimization problem in VQAs~\cite{2021_training_variational}.
By contrast, this paper aims to provide a heuristic optimizer that approximately solves this optimization problem with a small number of measurement shots.
In experiments using a quantum device, we can evaluate the cost function from the sum of the expectation values $\Tr[P_k U(\bm{\theta})(\Ket{0}\Bra{0})^{\otimes n} U^{\dag}(\bm{\theta})]$ for all $k$, each of which can be estimated by independently repeating the preparation of $U(\bm{\theta})\Ket{0}^{\otimes n}$ by the parameterized circuit and the measurement of this state in the eigenbasis of the Pauli operator $P_k$.
For each $k$, let $\overline{P}_k\in\mathbb{R}$ be a sample mean obtained from these measurements for $P_k$, and due to Eq.~\eqref{eq:observable}, we estimate $f(\bm{\theta})$ by
\begin{equation}
\label{eq:estimator}
    f(\bm{\theta})\approx\sum_k c_k\overline{P}_k.
\end{equation}
Each of these measurements is called a measurement shot.
In this way, we evaluate $f$ using a finite number of measurement shots;
in this setting, we are only allowed imprecise queries to the cost function due to statistical errors with the finite number of measurement shots.
Based on the central limit theorem~\cite{Kwak_Central}, we may model each imprecise query to $f(\bm{\theta})$ as
\begin{equation}
    y = f(\bm{\theta}) + \epsilon,
\label{eq: gaussian noise}
\end{equation}
where $y$ is an observed value, and $ \epsilon \sim \mathcal{N}(0, \sigma^2)$ is independent and identically distributed (IID) Gaussian noise.
From Hoeffding's inequality~\cite{doi:10.1080/01621459.1963.10500830}, to estimate $f(\bm{\theta})$ within an error $\epsilon$ with high probability, as large as $O(1/\epsilon^2)$ measurement shots may be required.
In practice, it is prohibitively costly (i.e., an excessive number of measurement shots are needed) to evaluate a well-approximated value of the cost function (as well as its gradient), which leads to significant overhead in performing VQAs~\cite{PhysRevA.92.042303,gonthier2020identifying}.

SGD aims to optimize a function $f(\bm{\theta})$ using an unbiased estimate of the gradient of $f$ to update the parameters $\bm{\theta}$ iteratively toward the optimal point with high probability.

In the optimization of circuit parameters for VQAs, we may need to evaluate the gradient of the cost function $f(\bm{\theta})$.
For $f(\bm{\theta})$ defined with parametric gates in the form of~\eqref{eq:parametric_gate},
we can utilize a parameter-shift rule~\cite{PhysRevA.98.032309,Schuld_2019} to calculate partial derivatives of the cost function from cost-function values at shifted circuit parameters, i.e.,
\begin{equation}
    \frac{\partial f(\bm{\theta})}{\partial \theta_i} = \frac{f(\bm{\theta}+\frac{\pi}{2}\bm{e}_i)-f(\bm{\theta} - \frac{\pi}{2}\bm{e}_i)}{2}.
\label{eq: shift-rule}
\end{equation}
Here $\theta_i$ is a circuit parameter allocated to the rotation angle of the $i$th Pauli rotation gate $U(\theta_i)=\exp(-\mathrm{i}P_i\theta_i)$, and $\bm{e}_i$ represents a unit vector along the coordinate of $\theta_i$.
Note that to obtain all the elements of the gradient of $f(\bm{\theta})$, we may need to evaluate each partial derivative independently.

However, as discussed above, we cannot exactly calculate the cost function and its gradient with a finite number of measurement shots, and the precise estimation of the gradient is costly in VQAs.
In this setting, a standard method for solving Eq.~\eqref{eq: problem statement minimization} is stochastic gradient descent (SGD)~\cite{10.1214/aoms/1177729586,Sweke2020stochasticgradient}, which updates the current point $\hat{\bm{\theta}}^{(t)}$ at iteration $t$ according to
\begin{equation}
    \hat{\bm{\theta}}^{(t+1)}=\hat{\bm{\theta}}^{(t)}-\eta^{(t)} \hat{\bm{g}}^{(t)}(\hat{\bm{\theta}}^{(t)}),
\label{eq: update rule of SGD}
\end{equation}
where $\eta^{(t)}$ is the step size,
and $\hat{\bm{g}}^{(t)}(\hat{\bm{\theta}}^{(t)})\coloneqq(\hat{g}^{(t)}_1(\hat{\bm{\theta}}^{(t)}),\ldots,\hat{g}^{(t)}_D(\hat{\bm{\theta}}^{(t)}))^\top$ is an unbiased estimator of the gradient $\nabla f(\hat{\bm{\theta}}^{(t)})$, i.e., $\mathbb{E}[\hat{\bm{g}}^{(t)}(\hat{\bm{\theta}}^{(t)})]=\nabla f(\hat{\bm{\theta}}^{(t)})$. Here $\hat{\bm{g}}^{(t)}$ is estimated with a finite number of measurement shots, i.e., with a shot size
\begin{equation}
\label{eq: shot size}
    \bm{s}_{\mathrm{grad}}^{(t)}=(s_1^{(t)},\ldots,s_D^{(t)})^\top.
\end{equation}
The estimate of each partial derivative is individually computed as
\begin{align}
\label{eq: g_i^t}
    \hat{g}^{(t)}_i(\bm{\theta})&=\frac{1}{s^{(t)}_{i}}\sum_{\mathsf{m}=1}^{s^{(t)}_{i}}g_{i}^{\mathsf{m}}(\bm{\theta}),\\
    g_{i}^{\mathsf{m}}(\bm{\theta}) &= (O_{+}^{\mathsf{m}}-O_{-}^{\mathsf{m}})/2,
\end{align}
where $O_{\pm}^{\mathsf{m}}$ is a single-shot estimator of $f(\bm{\theta}\pm\frac{\pi}{2}\bm{e}_i)$.
Each single-shot estimator of $f(\bm{\theta}\pm\frac{\pi}{2}\bm{e}_i)$ is constructed according to Eq.~\eqref{eq:estimator} by substituting $\bm{\theta}$ with $\bm{\theta}\pm\frac{\pi}{2}\bm{e}_i$, and the number of measurement shots used for estimating the $k$th term $c_k\overline{P}_k$ in Eq.~\eqref{eq:estimator} is denoted by $s_{i,k}^{(t)}$, which satisfies $\sum_{k}s_{i,k}^{(t)}=s_{i}^{(t)}$.
Given the shot size $\bm{s}_{\mathrm{grad}}^{(t)}$, each $s_{i,k}^{(t)}$ is probabilistically determined using a multinomial distribution in such a way that the probability $p_k$ of measuring the $k$th term should be proportional to the weight $|c_k|$, i.e., $p_k\propto|c_k|$ and $\sum_k p_k=1$~\cite{arrasmith2020operator}; that is, it should hold that $\mathbb{E}[s_{i,k}^{(t)}]=p_k s_{i}^{(t)}$ for each $k$ and $i$.
Since the gradient is estimated from two values $f(\bm{\theta}\pm\frac{\pi}{2}\bm{e}_i)$ of the cost function, the number of measurement shots used for obtaining $\hat{\bm{g}}^{(t)}$ is
\begin{equation}
    \sum_{i=1}^{D}2s_i^{(t)}=2s_{\mathrm{grad}}^{(t)},
    \label{eq:SGD_shot_size}
\end{equation}
where we write $s_{\mathrm{grad}}^{(t)}\coloneqq\|\bm{s}_{\mathrm{grad}}^{(t)}\|_1$.

The estimator $\hat{\bm{g}}^{(t)}(\bm{\theta})$ in VQAs is unbiased for all $\bm{\theta}\in\mathbb{R}^{D}$, which is a preferable property to achieve convergence of SGD~\cite{Sweke2020stochasticgradient}.
In addition, to guarantee convergence of SGD, we may require the step size to vanish as the estimated points approach a minimizer.
In this case, the SGD achieves the optimization to accuracy $\epsilon$ within  $O(1/\epsilon^4)$ iterations in general for non-convex functions~\cite{bottou2018optimization}, such as typical cost functions in VQAs. 
However, in practice, a user needs to designate a specific decay rate of step size to achieve good performance, whose optimization can be difficult.

BO is a gradient-free framework for optimization of an unknown function $f(\bm{\theta})$~\cite{7352306,article_practical_bayesian}. BO can be employed to optimize an expensive-to-evaluate cost function in settings where only noisy observations of the function are possible, and we try to seek a minimizer of $f(\bm{\theta})$ with as small a number of noisy observations as possible.
One of the features of BO is to utilize an easy-to-compute surrogate model that approximates the unknown cost function based on observed data~\cite{bodin2020modulating_ICML,NIPS2016_a96d3afe,snoek2015scalable}. 
A popular surrogate model for BO is Gaussian process (GP)~\cite{10.5555/1162254}.
GP is a collection of random variables such that every finite subset of random variables obeys a multivariate normal distribution.
In the BO, we put a GP prior over the true function $f(\bm{\theta})$ as $f(\bm{\theta}) \sim \mathcal{GP}(\mu(\bm{\theta}), k(\bm{\theta}, \bm{\theta}'))$, where $\mu(\bm{\theta})=\mathbb{E}(f(\bm{\theta}))$ is a mean function, $k(\bm{\theta}, \bm{\theta}')$ is a covariance kernel function. 
In practice, if one has no prior knowledge about the mean of the function $\mu(\bm{\theta})$ that one tries to fit, $\mu(\bm{\theta})$ can be set to $0$.
A major choice of the kernel function is a Gaussian kernel
\begin{equation}
    k(\bm{\theta}, \bm{\theta}') = \tau^2 \exp (\frac{-\|\bm{\theta}-\bm{\theta}'\|^2}{2l^2}),
\label{eq: SE kernel}
\end{equation}
where $\tau^2$ is called the signal variance that determines the average of the differences from the mean of the function, and $l$ is called the length scale that determines the length required for the values of the function to be uncorrelated~\cite{10.5555/1162254}.
For other conventional kernel functions, e.g., a Mat\'{e}rn kernel, see Ref.~\cite{10.5555/1162254}.

Here we consider a situation where we have a set of $N$ noisy observations of the cost function $\mathcal{D}_{1:N}=\{(\bm{\theta}^{(i)}, y^{(i)})\}_{i=1}^{N}$ at points $\bm{\theta}^{(1)},\ldots,\bm{\theta}^{(N)}$, where each $y^{(i)}= f(\bm{\theta}^{(i)})+\epsilon$ suffers from the IID Gaussian noise $ \epsilon \sim \mathcal{N}(0, \sigma^2)$.
Assuming that these observations are given according to GP\@,
we calculate a GP posterior conditioned on these estimations, which is governed by hyperparameters, namely, the signal variance $\tau^2$, the length-scale $l$, and the variance of Gaussian noise $\sigma^2$.
These hyperparameters can be estimated by means of maximizing a log marginal likelihood~\cite{10.5555/1162254}.
Then, if we observe the cost function $f$ at a new point $\bm{\theta}_*$, the value to be observed will obey a GP posterior expressed as
\begin{equation}
\begin{split}
    f_*|&\bm{\theta}_*, \mathcal{D}_{1:N}\\
    &\sim \mathcal{N}(\bm{k}_*^\top [K+\sigma^2 I]^{-1} \bm{y},  k_{**}-\bm{k}^{\top}[K+\sigma^2 I]^{-1}\bm{k}),
\label{eq: GP posterior}
\end{split}
\end{equation}
where $f_*=f(\bm{\theta}_*)$, $\bm{k}_*=[k(\bm{\theta}_*, \bm{\theta}^{(1)}),\cdots, k(\bm{\theta}_*, \bm{\theta}^{(N)})]^{\top}$, $k_{**}=k(\bm{\theta}_*,\bm{\theta}_*)$, and $K$ is the covariance matrix $[k(\bm{\theta}^{(i)}, \bm{\theta}^{(j)})]_{i,j=1}^{N}$~\cite{10.5555/1162254}.

In BO, we construct an acquisition function $\varphi(\bm{\theta})$ from the posterior in Eq.~\eqref{eq: GP posterior} and determine the next query point according to
\begin{equation}
\label{eq:min_acquisition}
    \bm{\theta}^{(N+1)}=\argmin_{\bm{\theta}\in \mathbb{R}^{D}}\varphi(\bm{\theta}).
\end{equation}
Several ways of constructing the acquisition function have been proposed, such as Thompson sampling~\cite{article_Thompson}, upper confidence bound~\cite{Srinivas_2012}, and expected improvement~\cite{article_A_Taxonomy}.
In particular, Thompson sampling estimates values of $f$ at a given set of points by sampling according to the multivariate normal distributions obtained from Eq.~\eqref{eq: GP posterior}, and use these sampled values as the values of the acquisition function at these points.
Then, we take the minimum among the values of the acquisition function for the set of points and perform the next query to $f$ at the minimum point in the set as shown in Eq.~\eqref{eq:min_acquisition}.
The minimization of $\varphi(\bm{\theta})$ is performed by using efficient optimization heuristics~\cite{Spall1998ANOO, Jones2001}.
BO proceeds with querying the cost function at the minimizer of $\varphi(\bm{\theta})$ and iteratively update the GP posterior according to Eq.~\eqref{eq: GP posterior} until a fixed number of queries to the cost function are performed~\cite{7352306}.

This framework of BO has been shown to reduce the required number of queries to the cost function in achieving the minimization compared to other global optimization algorithms~\cite{7352306}.
The performance of BO itself is governed by the ability to find the minimizer of $\varphi(\bm{\theta})$, which is also non-convex as well as the cost function.
Thus, it is important to design the acquisition function suitably so that the computational cost is relatively low and optimization heuristics are tractable~\cite{rolland2018high,pmlr-v37-kandasamy15,NIPS2013_8d34201a,kirschner2019adaptive_ICML}.
However, if the acquisition function is defined in a high-dimensional parameter space that typically appears in VQAs, it is excessively costly to use the BO\@.

\section{Results \label{sec: Our method}}
In the following, we present the description of SGLBO and introduce the adaptive shot strategy and suffix averaging.
Moreover, numerical experiments are provided to demonstrate the advantage of SGLBO compared to other state-of-the-art optimizers for VQAs\@.

\begin{figure}[t]
  \begin{algorithm}[H]
    \caption{Stochastic gradient line Bayesian optimization (SGLBO)}
    \label{alg1}
    \begin{algorithmic}[1]
      \Require Cost function $f(\bm{\theta})$ with $D$ parameters in Eq.~\eqref{eq: cost function}, the initial shot size $\bm{s}_{\mathrm{grad}}^{(0)}$ for evaluating the gradient in Eq.~\eqref{eq: shot size}, a kernel $k(\bm{\theta}, \bm{\theta}')$ and an acquisition function $\varphi(\bm{\theta})$ used for GP in Eqs.~\eqref{eq: GP posterior} and~\eqref{eq:min_acquisition}, the initial point $\hat{\bm{\theta}}^{(0)}$ to be updated according to Eq.~\eqref{eq: update rule of SGLBO}, the bound $\eta_{\max}$ of the $1$D subspace $\mathcal{L}^{(t)}$ to perform the BO in Eq.~\eqref{eq: 1D subspace}, the initial number $s_{\mathrm{cost}}^{(0)}$ of measurement shots for evaluating the cost function in BO in Eq.~\eqref{eq:s_cost}, the number $N=N_\mathrm{init}+N_\mathrm{eval}$ of queries used for the BO in Eq.~\eqref{eq:N}, the total number $s_{\rm{tot}}$ of measurement shots for the stopping condition~\eqref{eq:stopping_condition}, the precision $\kappa$ in estimating the gradient according to Eq.~\eqref{eq: the number of shot in SGLBO}, the description of the lower bound $G_\mathrm{grad}^{(t)}$ of the shot size in Eq.~\eqref{eq: the number of shot in SGLBO}, the description of the lower bound $G_\mathrm{cost}^{(t)}$ of the number of measurement shots in estimating the cost function in Eq.~\eqref{eq:s_cost_t}, a parameter $\alpha$ for suffix averaging in Eq.~\eqref{eq: suffix averaging}.
      \Initialize{$t\leftarrow0, \quad s_{\mathrm{temp}}^{(t)} \leftarrow 0$}
      \While {$s_{\mathrm{temp}}^{(t)}<s_{\rm{tot}}$} \Comment{Iterate until the stopping condition~\eqref{eq:stopping_condition} is satisfied.}
        \State $\hat{\bm{g}}^{(t)}, S^{(t)} \leftarrow $ Estimate the gradient $\mathbb{E}[\hat{\bm{g}}^{(t)}]=\nabla f(\hat{\bm{\theta}}^{(t)})$ using $2\times \bm{s}_{\mathrm{grad}}^{(t)}$ measurement shots according to Eq.~\eqref{eq: g_i^t}, and calculate its empirical variance $S^{(t)}$ in Eq.~\eqref{eq: the number of shot in SGLBO}.
        \State $\mathcal{L}^{(t)} \leftarrow$ Take the $1$D subspace $\mathcal{L}^{(t)}$ depending on $\hat{\bm{\theta}}^{(t)}, \hat{\bm{g}}^{(t)}, \eta_{\max}$ according to Eq.~\eqref{eq: 1D subspace}.
        \State $\hat{\bm{\theta}}^{(t+1)} \leftarrow$ Determine $\hat{\bm{\theta}}^{(t+1)}$ by the BO on $\mathcal{L}^{(t)}$ with $k(\bm{\theta}, \bm{\theta}'), \varphi(\bm{\theta}),N_\mathrm{init},N_\mathrm{eval}$ as described in the main text below Eq.~\eqref{eq:N}.
        \State $\bm{s}_{\mathrm{grad}}^{(t+1)} \leftarrow$ Determine the shot size for estimating the gradient, from $\kappa, \hat{\bm{g}}^{(t)}, S^{(t)}, D, G_\mathrm{grad}^{(t)}$ according to Eq.~\eqref{eq: the number of shot in SGLBO}.
        \State $s_{\mathrm{cost}}^{(t+1)} \leftarrow$ Determine the number of measurement shots for estimating the cost function in the BO\@, from $\bm{s}_{\mathrm{grad}}^{(t+1)}, G_\mathrm{cost}^{(t)}$ according to Eq.~\eqref{eq:s_cost_t}.
        \State $s_{\mathrm{temp}}^{(t+1)} \leftarrow  s_{\rm{temp}}^{(t)}+ 2s_\mathrm{grad}^{(t)}+Ns_{\rm{cost}}^{(t)}$ due to Eq.~\eqref{eq:iteration_shots}.
        \State $t \leftarrow t+1$
      \EndWhile
      \State $T \leftarrow t$
      \State \Return $\overline{\bm{\theta}}_{\alpha,T} \leftarrow$ Take the suffix average according to Eq.~\eqref{eq: suffix averaging}.
    \end{algorithmic}
  \end{algorithm}
\end{figure}

\subsection*{Description of algorithm \label{sec: description of algorithm}}

We present a framework for the optimizer of parameterized quantum circuits in the VQAs, stochastic gradient descent line Bayesian optimization (SGLBO). 
The idea behind SGLBO is to estimate the direction of the gradient based on SGD and further to utilize BO to estimate the optimal step size within the one-dimensional subspace of parameters in this direction.
This allows us to avoid the difficulty of choosing an appropriate step size in SGD, and also to achieve a feasible use of BO by limiting the domain to apply the BO to the one-dimensional space.
In addition, we introduce two noise-reduction techniques, adaptive shot strategy and suffix averaging, to improve the speed and the accuracy of minimizing the cost function.
Adaptive shot strategy and suffix averaging are crucial and characteristic components for the feasibility of SGLBO and will be explained in
``Adaptive shot strategy'' section and ``Suffix averaging for SGLBO'' section. 
Below, we will present the procedure of SGLBO (see also Algorithm~\ref{alg1}).

The SGLBO achieves the minimization of the cost function by iteratively updating the points to estimate the minimizer of the cost function. 
Let $T$ denote the total number of iterations in the SGLBO\@.
For each iteration $t=0,1,\ldots,T-1$,
let $\hat{\bm{\theta}}^{(t)}$ denote the point obtained in the $(t+1)$th iteration of the SGLBO, which is an estimator of the circuit parameters that minimize the cost function, and the initial point $\hat{\bm{\theta}}^{(0)}$ represents an initial guess of the minimizer.
Note that we here take $\hat{\bm{\theta}}^{(0)}$ uniformly at random, but in case a better initial guess of the minimizer than the uniformly random point is available, $\hat{\bm{\theta}}^{(0)}$ could be chosen as the better guess~\cite{Grant2019initialization,mitarai2020quadratic}.
Similarly to the SGD, the SGLBO computes an unbiased estimator $\hat{\bm{g}}^{(t)}$ of the gradient of the cost function at the point $\hat{\bm{\theta}}^{(t)}$, using $2s_{\mathrm{grad}}^{(t)}$ measurement shots due to Eq.~\eqref{eq:SGD_shot_size}.
The shot size $\bm{s}_{\mathrm{grad}}^{(t)}$ is determined in each iteration $t$ based on adaptive shot strategy, which will be explained in 
the following section.
Using $\hat{\bm{g}}^{(t)}$, the SGLBO updates the point $\hat{\bm{\theta}}^{(t)}$ to the next point according to an update rule described by
\begin{equation}
    \hat{\bm{\theta}}^{(t+1)}=\hat{\bm{\theta}}^{(t)}-\hat{\eta}^{*(t)}\hat{\bm{g}}^{(t)},
\label{eq: update rule of SGLBO}
\end{equation}
where $\hat{\eta}^{*(t)}$ is an estimator of the optimal step size.
The optimal step size $\eta^{*(t)}$ is defined as
\begin{equation}
    \bm{\theta}^{*(t)}\coloneqq \argmin_{\bm{\theta} \in \mathcal{L}^{(t)}}f(\bm{\theta})=\hat{\bm{\theta}}^{(t)} - \eta^{*(t)} \hat{\bm{g}}^{(t)},
\label{eq: optimal step size}
\end{equation}
where $\mathcal{L}^{(t)}$ is the one-dimensional subspace for applying the BO, i.e., 
\begin{equation}
    \mathcal{L}^{(t)} \coloneqq \{ \hat{\bm{\theta}}^{(t)} - \eta^{(t)} \hat{\bm{g}}^{(t)} \mid \eta^{(t)} \in [-\eta_{\rm{max}}, \eta_{\rm{max}}]\},
\label{eq: 1D subspace}
\end{equation}
and $\eta_{\rm{max}}>0$ is a constant hyperparameter to bound the one-dimensional subspace that will be specified in 
``Example of choice of hyperparameters and implementation'' section.
We remark that we choose $\eta_{\rm{max}}$ as a constant independent of $D$ so that the BO should be feasible even in the case of large $D$.
A parameter region of $D$ parameters of a circuit can be a $D$-dimensional hypercube, e.g., $\bm{\theta}\in[-\pi,\pi]^D$ for the circuit in Fig.~\ref{fig: ansatz}, 
and thus, to cross the whole parameter region by $\mathcal{L}^{(t)}$, one may be tempted to choose $\eta_{\rm{max}}$ as the length of the diagonal of this $D$-dimensional hypercube, i.e., $\eta_{\rm{max}}\approx \sqrt{D}$; however, for the feasibility of the BO, it is indeed essential to keep $\eta_{\rm{max}}$ constant.
Our approach can be considered an improvement over the SGD with a constant step size $\eta_{\rm{max}}$, where we use the BO to estimate the optimal step size $\hat{\eta}^{*(t)}$ instead of using the fixed step size $\eta_{\rm{max}}$.

To obtain an estimate of the optimal step size $\hat{\eta}^{*(t)}$ in Eq.~\eqref{eq: update rule of SGLBO}, we perform the procedure of BO on $\mathcal{L}^{(t)}$ by using a fixed number of measurement shots
\begin{equation}
\label{eq:s_cost}
    s_{\mathrm{cost}}^{(t)}
\end{equation}
per query to the cost function, and querying these noisy observations of the cost function $N$ times in total with
\begin{equation}
\label{eq:N}
    N=N_{\mathrm{init}}+N_{\mathrm{eval}}.
\end{equation}
where $N_{\mathrm{init}}$ is the number of points used for initial evaluation for BO, and $N_{\mathrm{eval}}$ is the number of points evaluated during the BO in each step in addition to $N_{\mathrm{init}}$. This procedure determines $\hat{\eta}^{*(t)}$ in such a way that $\hat{\bm{\theta}}^{(t+1)}$ in Eq.~\eqref{eq: update rule of SGLBO} should be given by $\bm{\theta}^{(N+1)}$ in Eq.~\eqref{eq:min_acquisition}.
We will specify $N_{\mathrm{init}}$ and $N_{\mathrm{eval}}$ in 
``Example of choice of hyperparameters and implementation'' section.
In the BO, we use $N_{\mathrm{init}}$ points for the initial queries, which we take at equal intervals in the $1$D subspace $\mathcal{L}^{(t)}$.
Using the observed points, the BO iterates a cycle according to Eqs.~\eqref{eq: GP posterior} and~\eqref{eq:min_acquisition} to decide an additional point to evaluate per cycle.
Repeating $N_{\mathrm{eval}}$ cycles, we have $N_{\mathrm{eval}}$ points in addition to the $N_{\mathrm{init}}$ initial points, where the $n$th cycle for $n\in\{1,\ldots,N_{\mathrm{eval}}\}$ uses $(N_{\mathrm{init}}+n-1)$ points to decide the $(N_{\mathrm{init}}+n)$th point.
These $N$ points are used for the update according to Eq.~\eqref{eq: update rule of SGLBO}, i.e., the calculation of $\hat{\eta}^{*(t)}$.

In this way, the SGLBO updates the point $\hat{\bm{\theta}}^{(t)}$ according to Eq.~\eqref{eq: update rule of SGLBO} until we consume a preset total number of measurement shots $s_\mathrm{tot}$, which we initially designate.
In particular, in the $(t+1)$th iteration for each $t=0,\ldots,T-1$, we use $2 s_\mathrm{grad}^{(t)}$ measurement shots for estimating the gradient according to Eq.~\eqref{eq:SGD_shot_size}, and also use $s_\mathrm{cost}^{(t)}$ measurement shots for each of the $N$ queries to the cost function in the BO\@; that is, the number of measurement shots that we use in the $(t+1)$th iteration is
\begin{equation}
\label{eq:iteration_shots}
    2s_\mathrm{grad}^{(t)}+Ns_\mathrm{cost}^{(t)}.
\end{equation}
In the SGLBO, if the total number of measurement shots used in the iterations exceeds the preset bound $s_\mathrm{tot}$, i.e.,
\begin{equation}
\label{eq:stopping_condition}
    \sum_{t=0}^{T-1}\left[2s_\mathrm{grad}^{(t)}+Ns_\mathrm{cost}^{(t)}\right]\geqq s_\mathrm{tot},
\end{equation}
then we stop the iterations.
Note that $T$ is given by the minimum number of iterations satisfying Eq.~\eqref{eq:stopping_condition}, determined during running the SGBLO depending on $s_\mathrm{tot}$.
We could also stop the iterations if we achieve the convergence of the cost function, while we here use the stopping condition based on $s_\mathrm{tot}$ for simplicity of presentation.
We remark that it would be too costly in VQAs to check the convergence of the values of the cost function $f(\hat{\bm{\theta}}^{(t)})$ itself, which we avoid here; instead, it would be possible, e.g., to use another stopping condition by checking the convergence of the sequence of parameters ${(\hat{\bm{\theta}}^{(t)})}_{t=0,\ldots,T-1}$.

Finally, after the last iteration, the optimizer calculates a suffix average~\cite{harvey2019tight} of the points ${(\hat{\bm{\theta}}^{(t)})}_{t=0,\ldots,T-1}$, i.e., an average of a subset of the points in a latter part of the iterations, which we will explain in
``Suffix averaging for SGLBO'' section.
This suffix average is output as the estimate of the minimizer of the cost function.

The procedure of the SGLBO may require an additional cost of measurement shots for the BO compared to the SGD without using the BO, but this cost is negligible as explained in the following.
To estimate the optimal step size by the BO, we may use an extra number of measurement shots to query the cost function, in addition to the gradient estimation based on the SGD\@.
For simplicity, suppose that the shot size~\eqref{eq: shot size} and the number of measurement shots to evaluate the cost function in the BO are given by a constant $s$, i.e., $s_i^{(t)}=s$ ($i\in\{1,\ldots,D\}$) and $s_{\mathrm{cost}}^{(t)}=s$.
Then, due to Eq.~\eqref{eq:iteration_shots}, the number of measurement shots to be used in each iteration of the SGLBO is
$(2D+N)s$.
In this case, the cost of estimating the optimal step size is the same as the cost of the gradient estimation for a parameterized quantum circuit with $N/2$ additional parameters.
This cost can be negligibly low as the number of circuit parameters $D$ gets large, and hence, we can indeed gain the benefit of estimating the optimal step size by the BO\@.

The foundation for why SGLBO can efficiently find a candidate of the minimum point, i.e., a stationary point, can be explained as follows.
The constant step-size SGD with averaging converges to a stationary point even in a non-convex setting~\cite{yu2020analysis}.
The SGLBO is designed to converge faster than this constant step-size SGD with averaging since we use the BO to find a step size that further reduces the value of the cost function compared to taking the deterministic constant step size.
In particular, in each step $t\in\{0,\ldots,T-1\}$, BO aims to find the minimum point along a $1$D subspace; that is, the cost function $f(\hat{\bm{\theta}}^{(t)})$ is reduced to $f(\hat{\bm{\theta}}^{(t+1)})$ satisfying $f(\hat{\bm{\theta}}^{(t+1)})\leqq f(\hat{\bm{\theta}}^{(t)})$ with high probability, in the case where BO is performed with sufficiently good precision.
In this case, as the iterations proceed, SGLBO improves the cost function according to $f(\hat{\bm{\theta}}^{(0)})\geqq f(\hat{\bm{\theta}}^{(1)})\geqq\cdots\geqq f(\hat{\bm{\theta}}^{(T-1)})$, which does not necessarily hold in SGD but should hold in the SGLBO with high probability, leading to an improvement compared to the mere use of the SGD\@.
We remark that the optimization problems in VQAs are non-convex, and hence, a tight analysis of the convergence speed would be challenging in general.
Some previous research such as Refs.~\cite{Sweke2020stochasticgradient, Harrow_2021} performs convergence analyses of optimizers for VQAs with assumptions on convexity or strong convexity, but the performance for non-convex problems that typically appear in VQAs are unknown.
In contrast, the above explanation of convergence does not require the convexity assumptions.
However, to bound the speed of convergence of SGLBO, further assumptions may be needed since non-convex optimization problems are hard to solve by nature.
We leave the tight analysis of the convergence speed of the SGLBO under an appropriate assumption for the setting of VQAs for further research; instead, we will use numerical simulation to show the fast convergence speed of the SGLBO in our numerical experiments.

\subsection*{\label{sec:adaptive_shot_strategy}Adaptive shot strategy}

The number of measurement shots used for estimating values and gradients of the cost function is one of the crucial parameters in stochastic optimization algorithms.
In such algorithms, we may have a trade-off between efficiency and accuracy. 
In particular, at the beginning of optimization, we can use an imprecise gradient estimated with few measurement shots to roughly move to points around the minimizer.
On the other hand, at the end of optimization, the gradients with less noise are needed to further decrease the value of the cost function. 
This observation motivates us to establish a strategy to gradually increase the shot size~\eqref{eq: shot size} used for estimating the gradient in the SGLBO as the optimization proceeds.

Such adaptive shot strategies have been well studied in the field of machine learning~\cite{Friedlander_2012,article_adaptive_sampling,article,article_Pasupathy,pmlr-v54-de17a,balles2017coupling,bollapragada2018a}, and one of them has been applied also in the context of VQAs~\cite{K_bler_2020,gu2021adaptive}. 
However, the formula for estimating the next number of measurement shots given in Refs.~\cite{K_bler_2020,gu2021adaptive} depends on the step size and becomes invalid when the step size exceeds a certain range.
Problematically, the step size in the SGLBO often exceeds the range.
Thus, our algorithm utilizes a different approach, the norm test~\cite{article_adaptive_sampling,pmlr-v54-de17a,article}, which determines the number of measurement shots to maintain a constant signal-to-noise ratio of the estimate of the gradient.

In the norm test, we want to decide the shot size based on a condition that the estimated vector $-\hat{\bm{g}}^{(t)}$ should be appropriately in a descent direction~\cite{pmlr-v54-de17a}, which ideally would be
\begin{equation}
    \delta^{(t)}\coloneqq\|\hat{\bm{g}}^{(t)}-\bm{\nabla} f(\hat{\bm{\theta}}^{(t)})\| \leqq \kappa \|\hat{\bm{g}}^{(t)}\|,
\label{Condition for estimates of gradient}
\end{equation}
with a parameter $\kappa$ satisfying $0\leqq \kappa<1$.
Intuitively, as the optimization proceeds, the norm $\|\hat{\bm{g}}^{(t)}\|$ of the gradient becomes small, and the condition~\eqref{Condition for estimates of gradient} requires that the estimate $\hat{\bm{g}}^{(t)}$ of the gradient should become precise as $\|\hat{\bm{g}}^{(t)}\|$ gets small.
However, the exact evaluation of $\delta^{(t)}$ would be prohibitively costly in VQAs\@.
Thus, we square both sides of the above inequality and then replace the left hand side with its expectation, i.e., $\mathbb{E}[(\delta^{(t)})^2]=\mathrm{Var}[\hat{\bm{g}}^{(t)}]$, where $\mathrm{Var}[\hat{\bm{g}}^{(t)}]$ is the variance of $\hat{\bm{g}}^{(t)}$.
The exact value of this variance is still difficult to calculate, and hence, we make the approximation using a sample variance~\cite{alma991005511429705181}, i.e.,
\begin{equation}
    \mathrm{Var}[\hat{\bm{g}}^{(t)}] \simeq \frac{\Tr(\Sigma^{(t)})}{s_{\mathrm{grad}}^{(t)}}
\end{equation}
where $\Sigma^{(t)}_{ij} \coloneqq \mathbb{E}[(g_i^{(t)} - \nabla f_i^{(t)})(g_j^{(t)} - \nabla f_j^{(t)}))]$.
Instead of Eq.~\eqref{Condition for estimates of gradient}, the norm test could check
\begin{equation}
    \frac{\Tr(\Sigma^{(t)})}{s_{\mathrm{grad}}^{(t)}}\leqq \kappa^2\|\hat{\bm{g}}^{(t)}\|^2.
\label{eq: condition for norm test: general}
\end{equation}
To adapt the condition~\eqref{eq: condition for norm test: general} to the setting of VQAs, we consider the freedom of choosing the number of measurement shots for estimating each partial derivative of the cost function in Eq.~\eqref{eq: shift-rule}.
Since each partial derivative is estimated independently, Eq.~\eqref{eq: condition for norm test: general} can be written as,
\begin{equation}
    \sum_{i}\frac{\left(\sigma_{i}^{(t)}\right)^2}{s^{(t)}_{i}}\leqq\kappa^2\|\hat{\bm{g}}^{(t)}\|^2,
\label{eq: condition for norm test: SGLBO}
\end{equation}
where $\sigma_i^{(t)}\coloneqq\sqrt{ \mathrm{Var}[g_i^{(t)}]}$.
Now we impose a constraint on the number of measurement shots so that each estimate of the partial derivative should have an equal variance, i.e., $(\sigma_{i}^{(t)})^2/s^{(t)}_{i}=(\sigma_{j}^{(t)})^2/s^{(t)}_{j}$ for $i\neq j$.
Then, we obtain a lower bound of $s_i^{(t)}$ for each $i$, i.e.,
\begin{equation}
    s_i^{(t)}\geqq \frac{1}{\kappa^2} \frac{\left(\sigma_{i}^{(t)}\right)^2D}{\|\hat{\bm{g}}^{(t)}\|^2}.
\label{eq: individual number of shots: SGLBO}
\end{equation}
In practice, the true variance $(\sigma_{i}^{(t)})^2$ is still too costly to evaluate, and thus, we replace it with the empirical variance $(S^{(t)})^2$, which is accessible.
Consequently, we forecast the number of measurement shots so that it should satisfy
\begin{equation}
    s_{i}^{(t+1)} \geqq \frac{1}{\kappa^2} \frac{\left(S_{i}^{(t)}\right)^2D}{\|\hat{\bm{g}}^{(t)}\|^2},
\end{equation}
which we use to estimate the gradient in the next iteration.
Since the SGLBO is intended to be applied to highly noisy cases, to avoid the cases where $s_{i}^{(t+1)}$ is too small to estimate the gradient appropriately, we here set a lower bound $G_\mathrm{grad}^{(t)}$ on the shot size and decide the next shot size according to
\begin{equation}
s_{i}^{(t+1)} =\max\left\{ \left\lceil\frac{1}{\kappa^2} \frac{\left(S_{i}^{(t)}\right)^2D}{\|\hat{\bm{g}}^{(t)}\|^2}\right\rceil,G_\mathrm{grad}^{(t)}\right\},
\label{eq: the number of shot in SGLBO}
\end{equation}
where $\lceil{}\cdots{}\rceil$ is the ceiling funciton.
The choice of $G_{\mathrm{grad}}^{(t)}$ will be specified in 
``Example of choice of hyperparameters and implementation'' section.

Using the shot size specified by Eq.~\eqref{eq: the number of shot in SGLBO}, we also decide the number of measurement shots used for observing values of the cost function in the BO according to
\begin{equation}
\label{eq:s_cost_t}
    s_{\mathrm{cost}}^{(t+1)}=\max \left\{\frac{1}{D}\sum_{i=1}^{D}s_{i}^{(t)}, G_{\mathrm{cost}}^{(t)} \right\},
\end{equation}
where $G_{\mathrm{cost}}^{(t)}>0$ is a constant for avoiding the cases where $s_{\mathrm{cost}}^{(t+1)}$ becomes too small to estimate the optimal step size appropriately.
The choice of $G_{\mathrm{cost}}^{(t)}$ will also be specified in
``Example of choice of hyperparameters and implementation'' section.

\subsection*{\label{sec:suffix_averaging}Suffix averaging for SGLBO}

In VQAs, one could use a point obtained from the final iteration as the result of the optimization.
However, in SGLBO, we use BO to estimate the optimal step size in Eq.~\eqref{eq: optimal step size}, and due to statistical error in the estimation, we suffer from the influence of the error between the estimate of the optimal step size obtained from the BO and the true optimal step size.
Moreover, hardware noise also prevents steady update of the points, especially when we use near-term noisy quantum devices.
Such errors or noises may lead to an oscillation of the points in the final part of the iterations around the minimizer. To suppress such oscillation, we take a suffix average of these points in the final part of the iterations, rather than using the single point of the final iteration itself.

Given the sequence of points obtained from $T$ iterations $\hat{\bm{\theta}}^{(0)}, \ldots, \hat{\bm{\theta}}^{(T-1)}$, the $\alpha$-suffix average is defined as the average of the last $\alpha T$ points~\cite{harvey2019tight}
\begin{equation}
    \overline{\bm{\theta}}_{\alpha, T} = \frac{1}{\alpha T}\sum_{t=(1-\alpha)T-1}^{T-1}\hat{\bm{\theta}}^{(t)},
\label{eq: suffix averaging}
\end{equation}
where $\alpha \in (0,1]$ is some constant, and $\alpha$ and $T$ are taken here in such a way that $\alpha T$ should be an integer.
During the optimization, we store the sequence of the points $(\hat{\bm{\theta}}^{(t)})_t$ in memory.
At the end of optimization, we calculate the suffix average of these points according to the above formula and output the suffix average as the result of the SGLBO\@.

Importantly, to achieve the goal of suppressing the effect of noise at the points in the final part of the iterations, the suffix averaging here uses an equal weight in averaging out the noise in this part.
To achieve this suppression with small overhead, the parameter $\alpha$ should be chosen appropriately, in such a way that the last $\alpha T$ points should be kept in a reasonably small fraction among all $T$ points yet still large enough to suppress the noise effectively.
We note that, instead of using the equal weight, averaging with a decaying sequence of weights would also work~\cite{shamir2013stochastic}, which may have a merit in a case where one does not have enough memory to store all points and wants to average the points on the fly.
Detailed comparison of suffix-averaging techniques using different sequences of weights in VQAs is left for future work.

The suffix averaging can accelerate the convergence of SGD in some cases; for example, for optimization of a strongly convex function, i.e., a function that is (roughly speaking) more convex than a quadratic function, the error of the point in the $T$th iteration decreases at the speed of $O(\log(T)/T)$ with high probability, but the error of the suffix average of the points in the latter half of the $T$ iterations reduces to $O(1/T)$, achieving the optimal speed~\cite{harvey2019tight}.
In the case of VQAs, $f$ may not be strongly convex.
However, even in the SGLBO, we can suppress the oscillation around the minimizer in practice by taking the suffix average, which contributes to improving the results of the optimization. 

\begin{figure*}[t]
    \centering
    \includegraphics[width=7.0in]{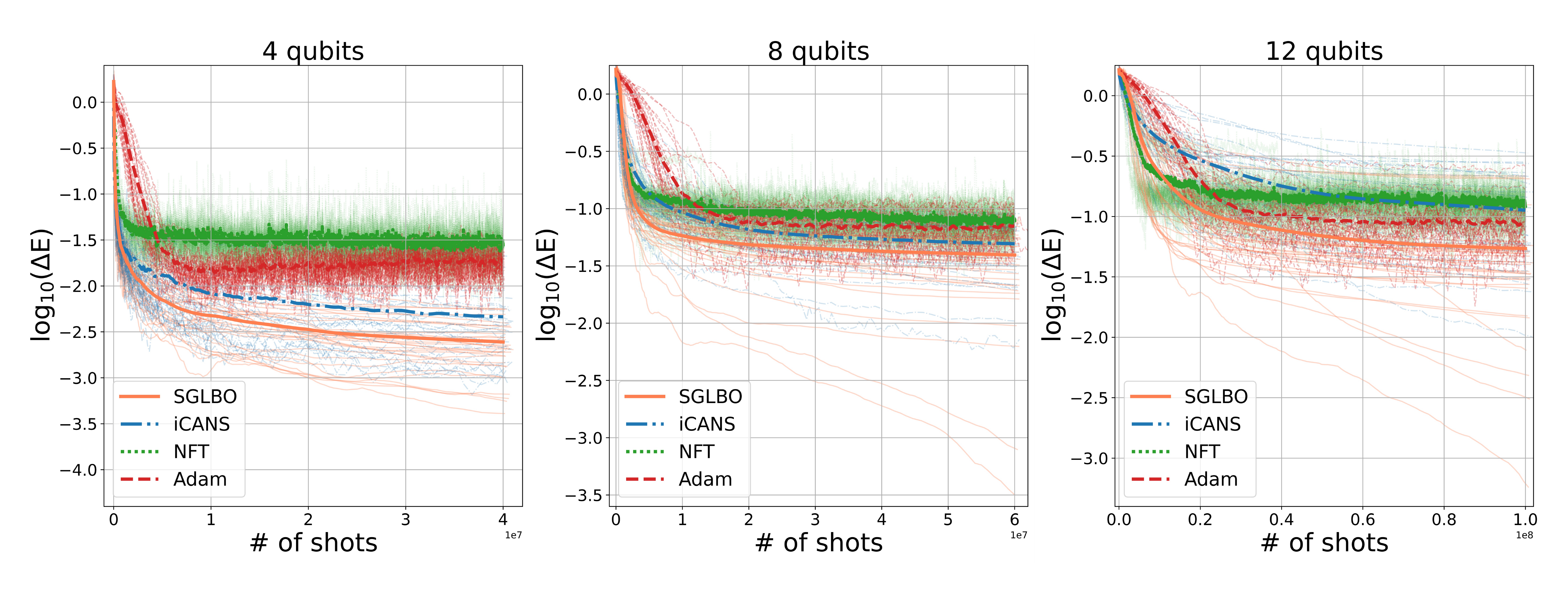}
    \caption{
    \textbf{Comparison of optimizers in terms of the performance on the VQE tasks.}
    We optimize the cost function of the VQE for $1$D transverse field Ising model with $n=4, 8, 12$ qubits in the noiseless case.
    In all plots, the $x$-axis represents the total number of measurement shots used during the optimization, and the $y$-axis represents the difference $\Delta E$ per site between the true value of the cost function (i.e., not evaluated with finite measurement shots) at each iteration and the minimum value of the cost function under the ansatz described in Fig.~\ref{fig: ansatz} with $r=4$. 
    For each optimizer, the thin lines represent each run repeated twice from fifteen different initial points, and the thick line represents the average of these thirty runs.
    Significantly, the SGLBO outperforms the other state-of-the-art optimizers in terms of both the convergence speed and the achievable accuracy for, a broad region $n=4, 8, 12$ of the number of qubits.}
    \label{fig: the results for VQE}
\end{figure*}

\subsection*{Example of choice of hyperparameters and implementation \label{sec: Example of choice of hyperparameters}}

We show an example of the choice of hyperparameters in Algorithm~\ref{alg1}.
These hyperparameters will be used in numerical experiments.
In the numerical experiments, we also consider the cases with and without hardware noise, referring to them as the noisy case and the noiseless case, respectively.

For estimating the gradient in the SGLBO,
we take the initial shot size as 
\begin{equation}
\label{eq:s_i_init}
    s_{i}^{(0)}=2\quad\text{for all $i$},
\end{equation}
and initialize $\hat{\bm{\theta}}^{(0)}$ by sampling from the uniform probability distribution.
We set the lower bound $G_\mathrm{grad}^{(t)}$ on the shot size by an average shot size in the last 10 iterations; i.e., for $t+1\geqq 10$, according to Eq.~\eqref{eq: the number of shot in SGLBO}, we take
\begin{align}
    &G_\mathrm{grad}^{(t)}=\frac{1}{10D}\sum_{i^\prime=1}^{D}\sum_{t^\prime=1}^{10}s_{i^\prime}^{(t-10+t^\prime)},\nonumber\\
    &\text{i.e.,}\;
    s_{i}^{(t+1)}=\nonumber\\
    &\max\left\{\left\lceil\frac{1}{\kappa^2} \frac{\left(S_{i}^{(t)}\right)^2D}{\|\hat{\bm{g}}^{(t)}\|^2}\right\rceil,\frac{1}{10D}\sum_{i^\prime=1}^{D}\sum_{t^\prime=1}^{10}s_{i^\prime}^{(t-10+t^\prime)}\right\},
\end{align}
and $G_\mathrm{grad}^{(t)}=1$ for $t\leqq 10$.
We set $\kappa=0.99$ in Eq.~\eqref{eq: the number of shot in SGLBO}.

In the BO that is used as a subroutine in the SGLBO,
we use the Gaussian kernel in Eq.~\eqref{eq: SE kernel} with $\tau^2=0.2$, $l=0.7$ as initial values. 
Before performing the GP regression to estimate values of a cost function, we optimize the hyperparameters, i.e., $\tau^2$, $l$, and the variance of Gaussian noise $\sigma^2$, by maximizing the marginal likelihood of the hyperparameters.
To avoid overfitting, we restrict the parameter region of these hyperparameters;
in our numerical experiments, we set the parameter region as $10^{-3}\leqq \tau^2 \leqq 5$, $10^{-3}\leqq l \leqq 1$, and $10^{-5}\leqq\sigma^2\leqq 5$.
In addition, we perform this hyperparameter optimization 10 times from uniformly random starting points and take the best parameters to ensure that the hyperparameters are not a poor local optimum.
As the acquisition function used in the BO, we choose Thompson sampling~\cite{article_Thompson,10.5555/1162254}.
After performing the BO, we set the estimated optimal step size as the minimum point of the predictive mean of a GP posterior conditioned on $N$ observed data points.

For the BO, we set $N_{\mathrm{init}}=5$ and $N_{\mathrm{eval}}=5$.
The $N_{\mathrm{init}}$ points of the initial evaluation is randomly chosen according to the uniform probability distribution over the $1$D subspace $\mathcal{L}^{(t)}$ in Eq.~\eqref{eq: 1D subspace} with 
\begin{equation}
    \eta^{(t)}\in[-\eta_{\mathrm{max}}, \eta_{\mathrm{max}}], \quad \eta_{\mathrm{max}} = \min\left\{\frac{\beta}{\|H\|}, \pi\right\},
\end{equation}
where $\|H\|$ is the operator norm, and $\beta>0$ is a constant that we set depending on the problem later in 
``Advantage of SGLBO for various system sizes'' section
and ``Robustness against hardware noise in SGLBO'' section.
Note that one of the initial evaluation points must be taken as $\eta^{(t)}=0$, i.e., the current point $\hat{\bm{\theta}}^{(t)}$, for the stability of the BO\@.
The number of measurement shots used for evaluating each point in the BO is given by Eq.~\eqref{eq:s_cost_t} with
\begin{align}
    &G_\mathrm{cost}^{(t)}=\frac{\|H\|^2}{\epsilon^2}~\text{for all}~t,\nonumber\\
    &\text{i.e.,}\;
    s_{\mathrm{cost}}^{(t)}=\max \left\{\frac{1}{D}\sum_{i=1}^{D}s_{i}^{(t)}, \frac{\|H\|^2}{\epsilon^2} \right\},
\end{align}
where $\epsilon=0.1$.
Given the outcomes of these measurements, we perform GP regression using GPy~\cite{GPy}. 

For the suffix averaging, we set $\alpha = 0.1$ in Eq.~\eqref{eq: suffix averaging}. 

\subsection*{Numerical experiments\label{sec: numerical experiments}}

In the following,
we numerically demonstrate the advantages of the SGLBO in comparison with state-of-the-art optimizers for VQAs\@.
The optimizers to be compared with the SGLBO are summarized in
``Optimizers for VQAs and their implementations'' section.
In particular, we investigate two situations: (1) when the size of a system scales up in
``Advantage of SGLBO for various system sizes'' section
, and (2) when hardware noise and connectivity between qubits on hardware are taken into account in
``Robustness against hardware noise in SGLBO'' section.
To this end, we simulate the performance of the optimizers in tasks of variational quantum eigensolver (VQE)~\cite{Peruzzo2014} for (1) and variational quantum compilation (VQC)~\cite{Khatri_2019} for (2).
Furthermore, we demonstrate in ``Merits of noise-reducing techniques for general optimizers'' section that the techniques of suffix averaging and adaptive shot strategy used in the SGLBO can also improve performance and noise robustness of a general class of optimizers, not only the SGLBO\@.

\subsection*{Optimizers for VQAs and their implementations
\label{sec: Implementations}}

To compare the SGLBO with other existing optimizers, we consider the following three state-of-the-art optimizers: adaptive moment estimation (Adam)~\cite{kingma2015adam}, individual coupled adaptive number of shots (iCANS)~\cite{K_bler_2020}, and Nakanishi-Fujii-Todo method (NFT)~\cite{PhysRevResearch.2.043158}.
Adam is a variant of SGD; although a number of different strategies for choosing step size in SGD have been proposed,  Adam chooses the step size adaptively based on the accumulated information of estimates of the gradient used in previous iterations.
The choice of step size in Adam is known to work well for many applications in the field of machine learning, but for VQAs, the required number of measurement shots for the optimization with Adam has been still prohibitively large~\cite{K_bler_2020}. We use Adam as a representative choice of a straightforward application of SGD to VQAs\@.
The iCANS is also a variant of stochastic gradient optimizers in which the number of measurement shots at each iteration is chosen frugally based on the first and second moment of the gradient to improve performance in VQAs\@.
While both of these optimizers are gradient-based optimizers, NFT is a sequential optimization method along an axis of the parameters using function fitting rather than the gradient. 

For iCANS, we in particular use iCANS1~\cite{K_bler_2020}, and for Adam, we used the same values of the hyperparameters as Ref.~\cite{K_bler_2020}. In terms of the initial number of measurement shots used in iCANS, which is not mentioned in Ref.~\cite{K_bler_2020}, we set $s_{i}^{(0)}=2$ for all $i$ in our numerical experiments.
Here we note that for iCANS1, the step size $\eta_t$ is changed depending on the tasks of VQAs as specified in 
``Advantage of SGLBO for various system sizes'' section and ``Robustness against hardware noise in SGLBO'' section, following Ref.~\cite{K_bler_2020}.
In addition, we used $s_{i}^{(t)}=1000$ shots for each evaluation of the cost function in Eq.~\eqref{eq: shift-rule} in Adam and $s_{\mathrm{cost}}^{(t)}=1000$ shots for each evaluation of the cost function to fit the function in NFT\@. Note that the values of the hyperparameters for which the optimizer works well are selected manually or by referring to the values of previous studies, and we did not perform an exhaustive hyperparameter search since such a search is computationally too costly to perform. After all, it may be infeasible to run such a hyperparameter search when we apply these optimizers to practical problems.

In these numerical experiments, we simulate quantum circuits by using Pennylane~\cite{bergholm2020pennylane}.
In ``Advantage of SGLBO for varisou system sizes'' section and
``Robustness against hardware noise in SGLBO'' section, the values of the cost function appearing in the figures are evaluated at the point of the final iterate in $(\hat{\bm{\theta}}^{(t)})_t$ (and the suffix averaged point in the SGLBO) by a noiseless simulator, where both the statistical noise and the hardware noise are ignored; in
``Merits of noise-reducing techniques for general
optimizers'' section,
these values are evaluated at the suffix averaged point by the noiseless simulator.
For each optimizer,
we repeated the overall optimization procedures fifteen times from uniformly random initial points, where each run from an initial point is repeated twice, and took the average over all the thirty runs.
In the figures, we display the logarithm of the average as a thick line and each run as a thin line, using log-linear plots.

\begin{figure*}[t]
    \centering
    \includegraphics[width=7.0in]{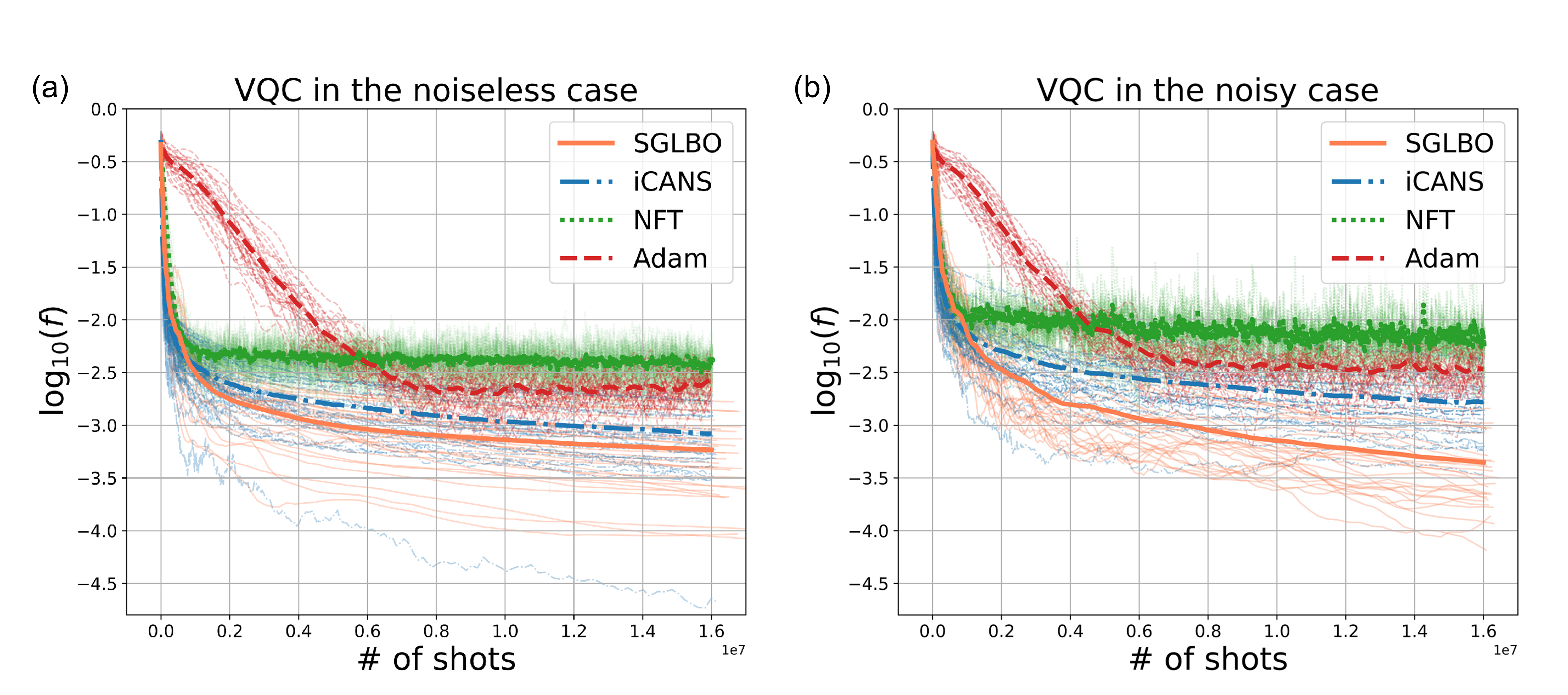}
    \caption{
    \textbf{Comparison of optimizers in terms of the performance on VQC tasks.}
    We optimize the cost function of the VQC task (a) without hardware noise and (b) with hardware noise for the ansatz circuit in Fig.~\ref{fig: ansatz} with $n=4$ qubits and $r=6$ repetitions.
    In both plots, $x$-axis represents the total number of measurement shots used during the optimization, and $y$-axis represents the cost-function value. 
    For each optimizer, the thin lines represent each run repeated twice from fifteen different initial points, and the thick line represents the average of these thirty runs.
    Remarkably, even under the moderate amount of the noise explained in the main text, the SGLBO can achieve almost the same accuracy as the noiseless case, whereas the achievable accuracy of the other state-of-the-art optimizers becomes worse in the noisy case.}
    \label{fig:the results for VQC}
\end{figure*}

\begin{figure*}[t]
    \centering
    \includegraphics[width=7.0in]{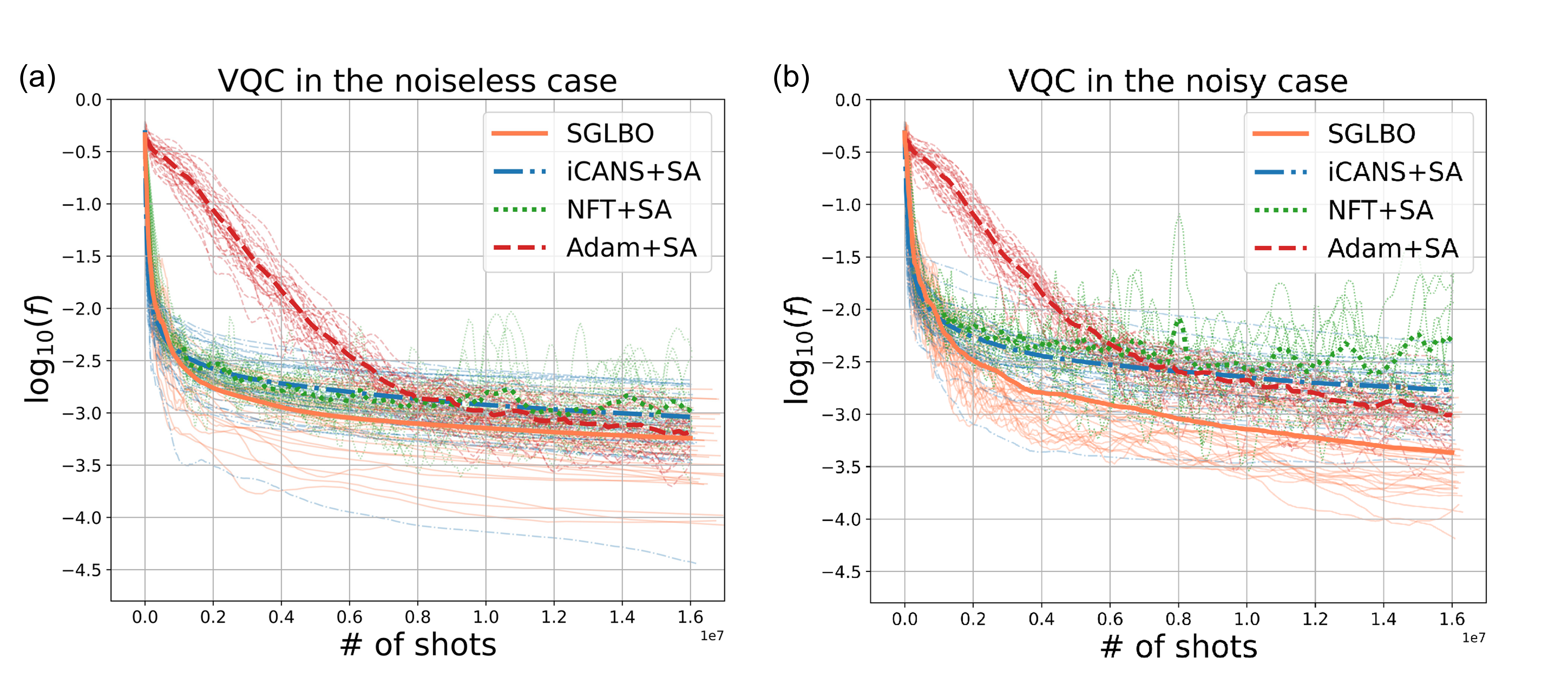}
    \caption{
    \textbf{Comparison of optimizers with the suffix averaging technique (SA), in the performance on the same VQC tasks as Fig.~\ref{fig:the results for VQC}.}
    The suffix averaging technique is not applied to iCANS, NFT, and Adam in Fig.~\ref{fig:the results for VQC} but is applied to all the optimizers in this figure.
    The $x$- and $y$-axes are the same as Fig.~\ref{fig:the results for VQC}.
    For each optimizer, the thin lines represent each run repeated twice from fifteen different initial points, and the thick line represents the average of these thirty runs.
    In both the noiseless (a) and noisy (b) cases, the technique of suffix averaging can significantly improve the accuracy of state-of-the-art optimizers, especially NFT and Adam, while the SGLBO still outperforms the others. This shows that the suffix averaging technique developed here is not only a particular technique for improving the SGLBO but can be a broadly applicable technique for designing an efficient optimizer for VQAs\@.}
    \label{fig:suffix}
\end{figure*}

\subsection*{Advantage of SGLBO for various system sizes}
\label{sec:VQE}

In this section, we investigate the performance of SGLBO as we scale up the system size.
We evaluate the performance of the optimizers in terms of the total number of measurement shots used during the optimization.
In each iteration, we calculate the difference per site between the cost-function value at the current point of each optimizer and the minimum value of the cost function.
In particular, we here consider a VQE task~\cite{Peruzzo2014} for a $1$D transverse field Ising model under open boundary conditions.
The VQE is an algorithm to calculate the ground state energy of a given Hamiltonian, where the cost function is defined as the expectation value of the Hamiltonian.
The Hamiltonian here is given by
\begin{equation}
    H= -J\left(\sum_{j=1}^{n-1} Z_j Z_{j+1} + g \sum_{j=1}^n X_{j}\right)
\end{equation}
where $Z_j$ and $X_j$ are the Pauli $Z$ and $X$ matrices, respecitvely, at the $j$th site on a $1$D chain of qubits, $J$ represents the energy scale, and $g$ is the relative strength of the external field compared to the nearest-neighbour couplings~\cite{PFEUTY197079}.
We choose $J=1.0$ and $g=1.5$.
We use the ansatz circuit in Fig.~\ref{fig: ansatz} with $r=4$ repetitions for $n=4, 8, 12$ qubits.
These sizes of the circuits are chosen based on the feasibility of classical simulation.
We remark that we do not change the depth of the ansatz circuits in this setting and change only the system size, so that the gradient does not vanish exponentially for the large system size~\cite{2018NatCo...9.4812M}; that is, it is expected that the problem of the barren plateau, which potentially make the optimization infeasible~\cite{2018NatCo...9.4812M,Cerezo_2021,PRXQuantum.2.040316}, is avoided in our setting.
In this problem, for the SGLBO, we restrict the region for the line search $\mathcal{L}_i$ by $\beta=3$, and for the iCANS, we set the step size $\eta_t=1/\|H\|$, following Ref.~\cite{K_bler_2020}.

The result of the numerical simulation is shown in Fig.~\ref{fig: the results for VQE}.
Significantly, we discover that the SGLBO outperforms the other optimizers~\cite{K_bler_2020, kingma2015adam,PhysRevResearch.2.043158} in all the cases of $n=4, 8, 12$ qubits, in terms of both the speed of convergence and the accuracy of estimating the minimum of the cost function. 
Thus, these advantages of the SGLBO can be obtained not only for the relatively small system size $n=4$ but more broadly for the larger system sizes $n=8,12$.
While NFT and Adam hit the limit of accuracy of the minimization in the early stage of the optimization, SGLBO and iCANS continue to improve the cost function even at the end of the optimization, which shows the advantage of deciding the number of measurement shots adaptively for each iteration in these algorithms.
Moreover, owing to using the BO for estimating the optimal step size in each iteration, the SGLBO enjoys faster convergence with a fewer number of overall measurement shots.
The additional cost of measurement shots in the BO in Eq.~\eqref{eq:iteration_shots} turns out to be negligible even on a small scale $n=4$, as well as the larger scales discussed in ``Description of algorithm'' section.
Consequently, for the VQE tasks in Fig.~\ref{fig: the results for VQE}, the SGLBO achieves the optimization of parameterized quantum circuits at the significantly faster convergence speed in terms of the number of measurement shots, and with better accuracy in minimizing the cost function than the other state-of-the-art optimizers.

\begin{figure*}[t]
    \centering
    \includegraphics[width=7.0in]{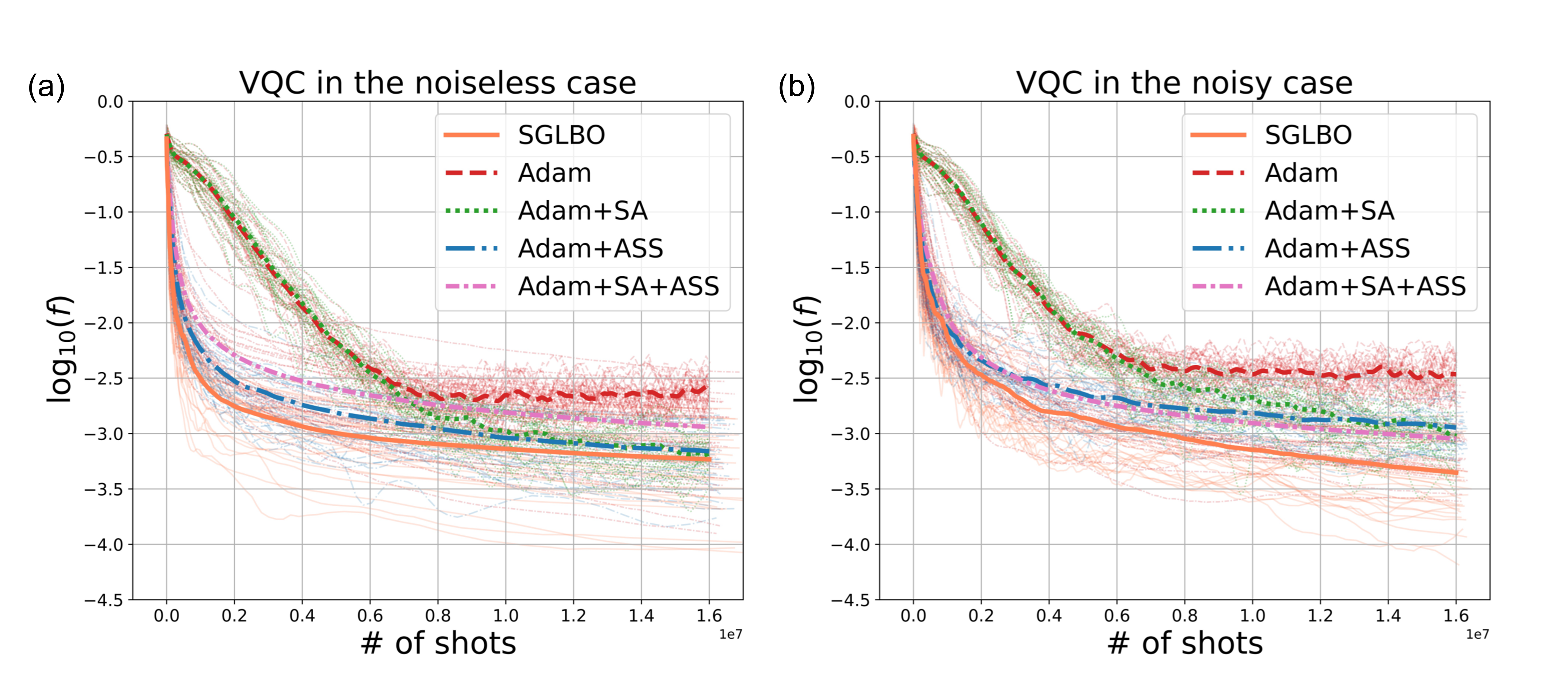}
    \caption{
    \textbf{Comparison of Adam with the suffix averaging technique (SA) and/or the adaptive shot strategy (ASS), in terms of the performance on the same VQC task as Fig.~\ref{fig:the results for VQC}.}
    The $x$- and $y$-axes are the same as Fig.~\ref{fig:the results for VQC}.
    For each optimizer, the thin lines represent each run repeated twice from fifteen different initial points, and the thick line represents the average of these thirty runs.
    In both the noiseless (a) and noisy (b) cases, the adaptive shot strategy improves the performance of the original Adam, but the SGLBO outperforms the others. This shows that the adaptive shot strategy is also useful in improving the accuracy of Adam, rather than a specific technique for the SGLBO\@.}
    \label{fig:the results of a variant of Adam for VQC}
\end{figure*}

\subsection*{Robustness against hardware noise in SGLBO}
\label{sec:VQC}

Next, we investigate the noise robustness of SGLBO\@.
We consider VQC~\cite{Khatri_2019} with a fixed input state.
The task of VQC is to find parameters of a parameterized circuit so that the unitary implemented by the circuit should act as equivalently as possible to a given target unitary when acting on a given input state.
Following Ref.~\cite{Khatri_2019},
we define the cost function as 
\begin{equation}
    f(\bm{\theta}) = 1- \frac{1}{n} \sum_{j=1}^{n} G^{(j)}_0,
\label{eq: cost function in VQC}
\end{equation}
where
\begin{equation}
\begin{split}
    &G^{(j)}_0 \\
    &= \Tr [(\dyad{0}{0}_j \otimes \mathbb{1}_{\Bar{j}})U^{\dag}(\bm{\theta})U(\bm{\theta}^*)(\dyad{0}{0})^{\otimes n}U^{\dag}(\bm{\theta}^*)U(\bm{\theta})].
\end{split}
\end{equation}
Here $\mathbb{1}_{\Bar{j}}$ is an identity operator acting on all qubits except the $j$th qubit, $G^{(j)}_0$ is the probability of getting the outcome $0$ on the $j$th qubit, $\bm{\theta}$ is a vector of circuit parameters to be optimized, and $\bm{\theta}^*$ is a target vector of circuit parameters that are chosen here as $\bm{\theta}^*=(0,\ldots,0)^\top\in\mathbb{R}^{D}$.
The target unitary is $U(\bm{\theta}^*)$, and the input state is $(\dyad{0}{0})^{\otimes n}$.
The ansatz circuit $U(\bm{\theta})$ used here is the one in Fig.~\ref{fig: ansatz} with $n=4$ and $r=6$.
In this case, the ansatz circuit can reach the optimal point at $\bm{\theta}=\bm{\theta}^*$ to output $(\dyad{0}{0})^{\otimes n}$, where the value of the cost function is exactly zero at the optimal point, and $y$-axis shows the difference between the true optimal value (i.e., zero) and the value at the estimated optimal point.
We note that this cost function is defined by local observables, so the gradient does not vanish in the shallow ansatz circuit used in this VQC task~\cite{Khatri_2019,Cerezo_2021}.
In VQC, we demonstrate the performance of the optimizers in both noiseless and noisy cases.
To simulate noise in the noisy case, we used information about the gate-operation and readout errors and the connectivity of IBM's Bogota processor~\cite{IBMQ2021,IBMQ2021Backend}. The detailed explanation on the parameters of the noise model is in Appendix~\ref{app: Error model}. We set $\beta=6$ to limit the region $\mathcal{L}_i$ for SGLBO and choose the step size $\eta_t=0.1$ for iCANS, following Ref.~\cite{K_bler_2020}.

The result of the numerical simulation is presented in Fig.~\ref{fig:the results for VQC}.
In the noiseless case, the SGLBO works better than the other state-of-the-art optimizers, which is consistent with the result of the VQE in Fig.~\ref{fig: the results for VQE}. 
Even more remarkably, even in the presence of a moderate amount of hardware noise described above, the SGLBO can achieve almost the same accuracy in minimizing the cost function as that in the noiseless case, while the other optimizers converge to worse cost-function values. 
This result indicates a remarkable noise resilience of the SGLBO, owing to using the BO and also the technique of suffix averaging.
In the SGLBO, the estimates of the minimizer of the cost function may be affected by hardware noise, and even if we use the BO that is relatively robust against the noise, these estimates may oscillate around the minimizer.
However, the suffix averaging of these estimates makes it possible to obtain a point that is even nearer to the minimizer.
In addition, the cost function in VQC has a preferable property that the minimizer is not susceptible to shifting caused by hardware noise~\cite{Sharma_2020}, and this property also contributes to the noise resilience in this case; that is, in other tasks for the VQAs without this property, the same accuracy as noiseless cases would be hard to achieve in noisy cases.
This result shows that the SGLBO can be more tolerant to hardware noise than the other state-of-the-art optimizers, which is crucial for the feasibility of performing VQAs on NISQ devices.

\subsection*{Merits of noise-reducing techniques for general optimizers}
\label{sec:suffix_averaging_demo}

We here also show that the technique of suffix averaging and adaptive shot strategy that we use in SGLBO turns out to be advantageous even in improving performance and noise robustness of the other state-of-the-art optimizers, not only the SGLBO\@.

In particular, we here consider the same task of VQC as ``Robustness against hardware noise in SGLBO'' section, and we first apply the suffix averaging technique to all the optimizers, i.e., iCANS, Adam, and NTF as well as SGLBO\@.
The result of the numerical simulation is shown in Fig.~\ref{fig:suffix}.
In both the noiseless and noisy cases, the technique of suffix averaging can significantly improve the accuracy of the state-of-the-art optimizers, especially NFT and Adam, compared to the cases without suffix averaging in Fig.~\ref{fig:the results for VQC}.
For iCANS, suffix averaging may not be as effective as NFT and Adam, but can still achieve a comparable accuracy to the cases without suffix averaging.
This result shows that the technique of suffix averaging that we apply in the SGLBO can indeed be useful as a general technique for improving a wide class of optimizers, not only for the SGLBO itself.
At the same time, our numerical simulation shows that even if we improve the other optimizers by the suffix averaging, the SGLBO still outperforms these optimizers.

Next, we apply the technique of adaptive shot strategy to Adam.
Note that our technique of adaptive shot strategy cannot be applied directly to NFT since NFT does not use gradient; also, iCANS uses its own variant of adaptive shot strategies, and hence, our technique based on the norm test cannot be combined with iCANS either without changing its own strategy.  
Following the setting of SGLBO with ~\eqref{eq:s_i_init}, we set $s_{i}^{(0)}=2$ for all $i$ when we combine the adaptive shot strategy with Adam in these experiments. The results of the numerical experiments are shown in Fig.~\ref{fig:the results of a variant of Adam for VQC}. In both noiseless and noisy cases, the adaptive shot strategy improves the performance of the original Adam. This indicates that the adaptive shot strategy based on the norm test is effectively applicable to the gradient-based optimizers and can improve the performance of the optimizers.
In Fig.~\ref{fig:the results of a variant of Adam for VQC}, we also demonstrate the combination of the suffix averaging and the adaptive shot strategy with Adam.
In noiseless case, since Adam with the adaptive shot strategy has not yet hit the floor in the minimization and is still improving its accuracy, taking suffix averaging worsened the accuracy, as opposed to the case of averaging out the noise around the optimal points.
On the other hand, in noisy case, the accuracy is improved.
This result further confirms the effectiveness of the suffix averaging technique against hardware noise.
The SGLBO still outperforms the other optimizers combined with these techniques.

In this way, the techniques that we develop for the SGLBO are also applicable broadly beyond the SGLBO itself, establishing a foundation for designing further efficient optimizers for VQAs in future research.
At the same time, these results show that SGLBO is an effective combination of all the techniques, i.e., SGD, BO, the suffix averaging, and the adaptive shot strategy, to outperform the state-of-the-art optimizers.

\section{Discussion \label{sec: Conclusion}}

In this work, we have developed an efficient framework, stochastic gradient line Bayesian optimization (SGLBO), for optimizing parameterized quantum circuits in variational quantum algorithms (VQAs)\@.
The core idea of the SGLBO is to estimate the direction of the gradient based on stochastic gradient descent (SGD), and also to use Bayesian optimization (BO) for estimating the optimal step size in this direction.
The BO used for estimating the optimal step size in the SGLBO contributes to minimizing the cost function faster and more accurately, owing to the robustness of the BO against noise.
To achieve the optimization feasibly within the fewer number of measurement shots,
we also formulated an adaptive measurement-shot strategy based on the norm test to estimate the direction of the gradient efficiently.
In addition, to suppress the effect of statistical error and hardware noise, we introduce the suffix averaging technique.
The SGLBO with these techniques can save the cost of the number of measurement shots in optimizing the parameterized circuits, and also improve the accuracy in minimizing the cost function in the VQAs\@.

To compare the performance of the SGLBO with other state-of-the-art optimizers, we numerically investigated two situations: (1) when the system size increases and (2) when the hardware noise is present. 
For various system sizes, we discover that the SGLBO significantly improves the required number of measurement shots for achieving a desired accuracy in minimizing cost functions, and reaches an even better accuracy in minimizing the cost functions than other state-of-the-art optimizers, as shown in Fig.~\ref{fig: the results for VQE}.
Furthermore, we have shown that, even in the presence of a moderate amount of hardware noise, the SGLBO can achieve almost the same accuracy as that in the noiseless case, whereas the accuracy of the other state-of-the-art optimizers has got worse, in the task shown in Fig.~\ref{fig:the results for VQC}.
To suppress the noise, the suffix averaging technique as well as the use of the BO is crucial, and it turns out that the suffix averaging and the adaptive shot strategy developed for the SGLBO can also improve the accuracy and the noise robustness of other existing optimizers as demonstrated in Fig.~\ref{fig:suffix}.

Consequently, integrating two different optimization approaches, SGD and BO, our results on the SGLBO open an alternative way to drastically reduce the cost of measurement shots in the optimization of parameterized quantum circuits, and also to make VQAs more feasible under unavoidable hardware noise in near-term quantum devices.
The techniques introduced here are versatile for problems with various system sizes, effective even in presence of noise, and widely applicable to a variety of algorithms for optimizing parameterized quantum circuits in the setting of VQAs, as demonstrated above.
At the same time, the approach developed for the SGLBO provides a fundamental insight into how VQAs can use classical information extracted from quantum states, progressing beyond estimating expectation values.
Moreover, the idea of the SGLBO indeed provides a general framework for optimizing noisy functions in the field of machine learning (ML), not specifically to VQAs.
Thus, our results are expected to be of interest not only to users of noisy intermediate-scale quantum (NISQ) devices but to much broader communities of quantum information, such as those working on ML-assisted calibration of quantum devices in experiments, quantum tomography using an ansatz, and quantum metrology.

These results point toward various directions of future research.
One possible direction is to investigate the difference in performance when the $1$D subspace for the BO currently taken in the gradient descent direction (Eq.~\eqref{eq: 1D subspace}) is chosen in another direction, such as natural gradient descent~\cite{Stokes_2020,Wierichs_2020,PRXQuantum.2.030324,haug2021optimal}, negative curvature descent~\cite{NEURIPS2018_f52854cc}, and conjugate gradient~\cite{10.1093/comjnl/7.2.149}.
Also, the development of a more efficient method for determining appropriate hyperparameter values in the SGLBO is also important for improving the accuracy.
In our work, we have empirically found that the SGLBO with suffix averaging performs well in practice even if hardware noise is considered,
but further research is needed to clarify of what class of hardware noise the suffix averaging can be tolerant, and how many iterations are needed to achieve comparable performance to the noiseless case.
It would also be interesting to provide a theoretical guarantee on the performance of the SGLBO under appropriate assumptions, especially in the setting of non-convex optimization\@; after all, both empirical and theoretical studies are crucial for harnessing the potential for near-term applications of VQAs.
Finally, since the SGLBO discovers a way to avoid the cost of precise estimation of expectation values in optimizing parameterized circuits for VQAs, it is even more advantageous to pursue applications of VQAs that do not require estimating the expectation values throughout running the entire algorithm, i.e., even after the optimization; for example, state-of-the-art quantum algorithms for quantum machine learning avoid the expectation-value estimation by solving sampling problems so that the speedup should not be canceled out~\cite{NEURIPS2020_9ddb9dd5,yamasaki2021exponential,kerenidis_et_al:LIPIcs:2017:8154}, and further research is needed to clarify how we can similarly avoid the expectation-value estimation in quantum machine learning with VQAs\@.\\

\section*{Data availability}
Data for the plots supporting the results in this work can be obtained from the corresponding author upon reasonable request.

\section*{Code availability}
Computer codes to perform the numerical experiments in this work are available from the corresponding author upon reasonable request.

\section*{Acknowledgment}
This work was supported by JST [Moonshot R\&D][Grant Number JPMJMS2061], JSPS Overseas Research Fellowships, and JST PRESTO Grant Number JPMJPR201A\@.

\section*{Author contributions}
S.T and H.Y. contributed to the initial conception of the ideas, to the
working out of details, and to the writing and editing of the manuscript.

\section*{Competing interests}
The authors declare no competing interests.

\appendix

\section{Error model\label{app: Error model}}
\begin{figure}
\includegraphics[width=3.4in]{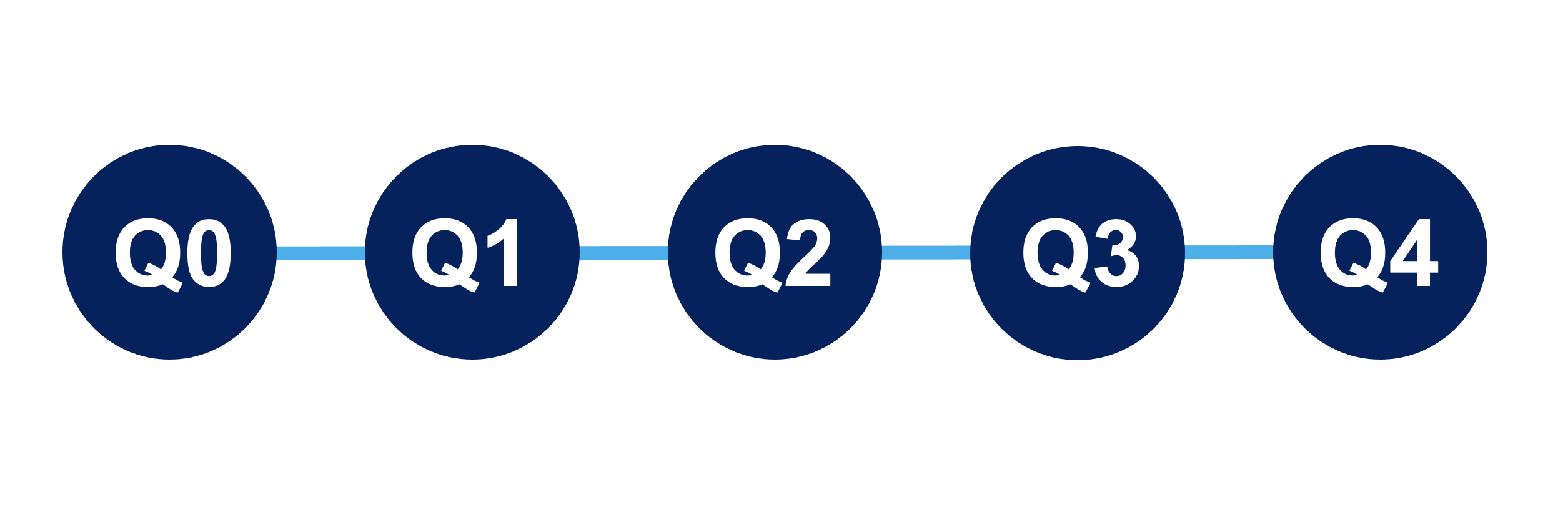}
\caption{The arrangement of qubits and the connectivity in {\tt{imbq\_bogota}}. The nodes represent qubits where each label corresponds to the noise profile shown in Table.~\ref{Tab: bogota noise profile}. The edges represent couplers, where we can apply a \textsc{CNOT} gate to the adjacent qubits connected by a coupler.}
\label{fig: bogota} 
\end{figure}

We explain the details of our numerical experiments in the noisy cases of the task of variational quantum compilation (VQC) presented in Sec.~IV~C of the main text.
To simulate noise, we use IBM's noisy simulator.
The simulator works with the basis gates, the noise profile, and the connectivity of IBM's quantum device, {\tt{imbq\_bogota}}~\cite{IBMQ2021, IBMQ2021Backend}.
The {\tt{imbq\_bogota}} has the following basis gates: \textsc{CNOT} gate, identity gate (ID), $X$ gate, $\sqrt{X}$ gate ($\pi/2$ rotation around $X$ axis), and $R_Z$ gate (arbitrary rotation around $Z$ axis).
Quantum circuits are decomposed into these basis gates using a default compiler provided by Qiskit~\cite{Qiskit}.
While we can apply the single-qubit operations to arbitrary qubits, the \textsc{CNOT} gate is allowed only between two adjacent qubits connected by a coupler, where the connectivity of couplers on {\tt{imbq\_bogota}} is shown in Fig~\ref{fig: bogota}.
In IBM's noisy simulator, one-qubit and two-qubit errors are modeled as a depolarizing error, and a measurement error is modeled as a single-qubit readout error for every measurement~\cite{IBMQ2021Noise}. 
The error parameters of one-qubit, two-qubit, and measurement errors are specified for each qubit as the gate error and the readout error, as shown in Table~\ref{Tab: bogota noise profile}.
We note that the $R_Z$ gate are implemented with no (negligible) error and zero (negligible) gate time in IBM's noisy simulator.

\bibliography{bibliography}

\begin{table*}[b]
\centering
\begin{tabular}{c||l|l||l|l|l}
 & ID error & $\sqrt{X}$ error & Single-qubit Pauli-X error & \textsc{CNOT} error & Readout assignment error  \\ \hline \hline
Q0  & $1.775\mathrm{e}$-4 & $1.775\mathrm{e}$-4   & $1.775\mathrm{e}$-4   & \begin{tabular}{c}
$0\_1$: $8.622\mathrm{e}$-3
\end{tabular}  & $1.58\mathrm{e}$-2   
\\ \hline
Q1  & $2.179\mathrm{e}$-4 & $2.179\mathrm{e}$-4   & $2.179\mathrm{e}$-4   &  \begin{tabular}{c}
$1\_2$: $7.100\mathrm{e}$-3\\ $1\_0$: $8.622\mathrm{e}$-3
\end{tabular} & $2.27\mathrm{e}$-2   \\\hline
Q2   & $2.005\mathrm{e}$-4 & $2.005\mathrm{e}$-4   & $2.005\mathrm{e}$-4   &  \begin{tabular}{c}
$2\_3$: $1.008\mathrm{e}$-2\\ $2\_1$: $7.100\mathrm{e}$-3
\end{tabular}& $1.51\mathrm{e}$-2   \\ \hline
Q3   & $7.687\mathrm{e}$-4 & $7.687\mathrm{e}$-4   & $7.687\mathrm{e}$-4   &  \begin{tabular}{c}
$3\_4$: $7.044\mathrm{e}$-3\\$3\_2$: $1.008\mathrm{e}$-2
\end{tabular}& $1.044\mathrm{e}$-1  \\ \hline
Q4   & $1.581\mathrm{e}$-4 & $1.581\mathrm{e}$-4   & $1.581\mathrm{e}$-4    &  \begin{tabular}{c}
$4\_3$: $7.044\mathrm{e}$-3
\end{tabular} & $2.48\mathrm{e}$-2   \\ \hline
\end{tabular}
\caption{Noise parameters of {\tt{imbq\_bogota}} used in our numerical simulation of VQC tasks in the noisy cases. In the columns of \textsc{CNOT} error, we write $i\_j$ to show noise information on a control qubit Q$i$ and a target qubit Q$j$ for $i,j\in\{0,1,2,3,4\}$.}
\label{Tab: bogota noise profile}
\end{table*}
\end{document}